\documentclass[aps,prd,twocolumn,superscriptaddress,preprintnumbers,notitlepage]{revtex4-1}

\usepackage{graphicx}
\usepackage{tensor}
\usepackage{multirow}
\usepackage{xcolor}
\usepackage{amsmath}
\usepackage{mathtools}

\graphicspath{{images/}{tikz/}}

\newcommand{\err}[2]{\tensor*[^{+#1}_{-#2}]{}{}}

\newcommand{\elp}{e^+}
\newcommand{\elm}{e^-}
\newcommand{\pip}{\pi^+}
\newcommand{\pim}{\pi^-}
\newcommand{\piz}{\pi^0}
\newcommand{\kp}{K^+}
\newcommand{\km}{K^-}
\newcommand{\ks}{K^0_S}
\newcommand{\pr}{p}
\newcommand{\apr}{\bar{p}}
\newcommand{\lam}{\Lambda}
\newcommand{\alam}{\bar{\Lambda}}

\newcommand{\Dz}{D^0}

\newcommand{\Bp}{B^+}
\newcommand{\Bm}{B^-}
\newcommand{\Bz}{B^0}

\newcommand{\decaykp}{B^+ \to h_c \kp}
\newcommand{\decayks}{B^0 \to h_c \ks}
\newcommand{\decaycckp}{B^+ \to (c \bar{c}) \kp}
\newcommand{\decayccks}{B^0 \to (c \bar{c}) \ks}

\newcommand{\invfb}{\text{fb}^{-1}}
\newcommand{\cm}{\text{cm}}
\newcommand{\mev}{\text{MeV}}
\newcommand{\mevcc}{\text{MeV}/c^2}
\newcommand{\gev}{\text{GeV}}

\newcommand{\gevcc}{\text{GeV}/c^2}

\newcommand{\mbc}{M_{\text{bc}}}
\newcommand{\de}{\Delta E}

\newcommand{\etaccha}[1]{
\ifcase#1\relax
\kp \ks \pim
\or
\kp \km \piz
\or
\ks \ks \piz
\or
\kp \km \eta_{2 \gamma}
\or
\kp \km \eta_{3 \pi}
\or
\kp \km \kp \km
\or
\kp \km \pip \pim
\or
\kp \km \pip \pim \piz
\or
\ks \km \pip \pim \pip
\or
\eta' (\to \eta_{2 \gamma} \pip \pim) \pip \pim
\or
\eta' (\to \eta_{3 \pi} \pip \pim) \pip \pim 
\or
\eta_{2 \gamma} \pip \pim
\or
\eta_{3 \pi} \pip \pim
\or
\pr \apr
\or
\pr \apr \piz
\or
\pr \apr \pip \pim
\or
\lam \alam
\or
\hccha{1}
\fi
}
\newcommand{\etacdec}[1]{\eta_c \to \etaccha{#1}}
\newcommand{\hccha}[1]{
\ifcase#1\relax
\eta_c \gamma
\or
\pr \apr \pip \pim
\fi
}
\newcommand{\hcdec}[1]{h_c \to \hccha{#1}}
\newcounter{optchannel}
\newcommand{\optcha}[1]{
\ifnum#1<17
\eta_c (\to \etaccha{#1}) \gamma
\else
{
\setcounter{optchannel}{#1}
\addtocounter{optchannel}{-16}
\hccha{\arabic{optchannel}}
}
\fi
}
\newcommand{\optdec}[1]{h_c \to \optcha{#1}}
\newcommand{\decayccchakp}[1]{B^+ \to (c \bar{c}) (\to \hccha{#1}) \kp}
\newcommand{\decayccchaks}[1]{B^0 \to (c \bar{c}) (\to \hccha{#1}) \ks}

\newcommand{\br}{\mathcal{B}}

\begin{document}

\title{Evidence for $\decaykp$ and observation of $\eta_c(2S) \to \hccha{1}$}

\noaffiliation
\affiliation{University of the Basque Country UPV/EHU, 48080 Bilbao}
\affiliation{Beihang University, Beijing 100191}
\affiliation{Brookhaven National Laboratory, Upton, New York 11973}
\affiliation{Budker Institute of Nuclear Physics SB RAS, Novosibirsk 630090}
\affiliation{Faculty of Mathematics and Physics, Charles University, 121 16 Prague}
\affiliation{Chonnam National University, Kwangju 660-701}
\affiliation{University of Cincinnati, Cincinnati, Ohio 45221}
\affiliation{Deutsches Elektronen--Synchrotron, 22607 Hamburg}
\affiliation{University of Florida, Gainesville, Florida 32611}
\affiliation{Key Laboratory of Nuclear Physics and Ion-beam Application (MOE) and Institute of Modern Physics, Fudan University, Shanghai 200443}
\affiliation{Justus-Liebig-Universit\"at Gie\ss{}en, 35392 Gie\ss{}en}
\affiliation{II. Physikalisches Institut, Georg-August-Universit\"at G\"ottingen, 37073 G\"ottingen}
\affiliation{SOKENDAI (The Graduate University for Advanced Studies), Hayama 240-0193}
\affiliation{Gyeongsang National University, Chinju 660-701}
\affiliation{Hanyang University, Seoul 133-791}
\affiliation{University of Hawaii, Honolulu, Hawaii 96822}
\affiliation{High Energy Accelerator Research Organization (KEK), Tsukuba 305-0801}
\affiliation{J-PARC Branch, KEK Theory Center, High Energy Accelerator Research Organization (KEK), Tsukuba 305-0801}
\affiliation{IKERBASQUE, Basque Foundation for Science, 48013 Bilbao}
\affiliation{Indian Institute of Science Education and Research Mohali, SAS Nagar, 140306}
\affiliation{Indian Institute of Technology Guwahati, Assam 781039}
\affiliation{Indian Institute of Technology Hyderabad, Telangana 502285}
\affiliation{Indian Institute of Technology Madras, Chennai 600036}
\affiliation{Indiana University, Bloomington, Indiana 47408}
\affiliation{Institute of High Energy Physics, Chinese Academy of Sciences, Beijing 100049}
\affiliation{Institute of High Energy Physics, Vienna 1050}
\affiliation{Institute for High Energy Physics, Protvino 142281}
\affiliation{INFN - Sezione di Napoli, 80126 Napoli}
\affiliation{INFN - Sezione di Torino, 10125 Torino}
\affiliation{Advanced Science Research Center, Japan Atomic Energy Agency, Naka 319-1195}
\affiliation{J. Stefan Institute, 1000 Ljubljana}
\affiliation{Institut f\"ur Experimentelle Teilchenphysik, Karlsruher Institut f\"ur Technologie, 76131 Karlsruhe}
\affiliation{Kennesaw State University, Kennesaw, Georgia 30144}
\affiliation{King Abdulaziz City for Science and Technology, Riyadh 11442}
\affiliation{Korea Institute of Science and Technology Information, Daejeon 305-806}
\affiliation{Korea University, Seoul 136-713}
\affiliation{Kyoto University, Kyoto 606-8502}
\affiliation{Kyungpook National University, Daegu 702-701}
\affiliation{LAL, Univ. Paris-Sud, CNRS/IN2P3, Universit\'{e} Paris-Saclay, Orsay}
\affiliation{\'Ecole Polytechnique F\'ed\'erale de Lausanne (EPFL), Lausanne 1015}
\affiliation{P.N. Lebedev Physical Institute of the Russian Academy of Sciences, Moscow 119991}
\affiliation{Ludwig Maximilians University, 80539 Munich}
\affiliation{Luther College, Decorah, Iowa 52101}
\affiliation{Malaviya National Institute of Technology Jaipur, Jaipur 302017}
\affiliation{University of Malaya, 50603 Kuala Lumpur}
\affiliation{University of Maribor, 2000 Maribor}
\affiliation{Max-Planck-Institut f\"ur Physik, 80805 M\"unchen}
\affiliation{School of Physics, University of Melbourne, Victoria 3010}
\affiliation{University of Mississippi, University, Mississippi 38677}
\affiliation{University of Miyazaki, Miyazaki 889-2192}
\affiliation{Moscow Physical Engineering Institute, Moscow 115409}
\affiliation{Moscow Institute of Physics and Technology, Moscow Region 141700}
\affiliation{Graduate School of Science, Nagoya University, Nagoya 464-8602}
\affiliation{Universit\`{a} di Napoli Federico II, 80055 Napoli}
\affiliation{Nara Women's University, Nara 630-8506}
\affiliation{National Central University, Chung-li 32054}
\affiliation{National United University, Miao Li 36003}
\affiliation{Department of Physics, National Taiwan University, Taipei 10617}
\affiliation{H. Niewodniczanski Institute of Nuclear Physics, Krakow 31-342}
\affiliation{Nippon Dental University, Niigata 951-8580}
\affiliation{Niigata University, Niigata 950-2181}
\affiliation{Novosibirsk State University, Novosibirsk 630090}
\affiliation{Osaka City University, Osaka 558-8585}
\affiliation{Pacific Northwest National Laboratory, Richland, Washington 99352}
\affiliation{Panjab University, Chandigarh 160014}
\affiliation{Peking University, Beijing 100871}
\affiliation{University of Pittsburgh, Pittsburgh, Pennsylvania 15260}
\affiliation{Punjab Agricultural University, Ludhiana 141004}
\affiliation{Theoretical Research Division, Nishina Center, RIKEN, Saitama 351-0198}
\affiliation{University of Science and Technology of China, Hefei 230026}
\affiliation{Seoul National University, Seoul 151-742}
\affiliation{Showa Pharmaceutical University, Tokyo 194-8543}
\affiliation{Soongsil University, Seoul 156-743}
\affiliation{Stefan Meyer Institute for Subatomic Physics, Vienna 1090}
\affiliation{Sungkyunkwan University, Suwon 440-746}
\affiliation{School of Physics, University of Sydney, New South Wales 2006}
\affiliation{Department of Physics, Faculty of Science, University of Tabuk, Tabuk 71451}
\affiliation{Tata Institute of Fundamental Research, Mumbai 400005}
\affiliation{Department of Physics, Technische Universit\"at M\"unchen, 85748 Garching}
\affiliation{Department of Physics, Tohoku University, Sendai 980-8578}
\affiliation{Earthquake Research Institute, University of Tokyo, Tokyo 113-0032}
\affiliation{Department of Physics, University of Tokyo, Tokyo 113-0033}
\affiliation{Tokyo Institute of Technology, Tokyo 152-8550}
\affiliation{Tokyo Metropolitan University, Tokyo 192-0397}
\affiliation{Virginia Polytechnic Institute and State University, Blacksburg, Virginia 24061}
\affiliation{Wayne State University, Detroit, Michigan 48202}
\affiliation{Yamagata University, Yamagata 990-8560}
\affiliation{Yonsei University, Seoul 120-749}
  \author{K.~Chilikin}\affiliation{P.N. Lebedev Physical Institute of the Russian Academy of Sciences, Moscow 119991} 
  \author{I.~Adachi}\affiliation{High Energy Accelerator Research Organization (KEK), Tsukuba 305-0801}\affiliation{SOKENDAI (The Graduate University for Advanced Studies), Hayama 240-0193} 
  \author{D.~M.~Asner}\affiliation{Brookhaven National Laboratory, Upton, New York 11973} 
  \author{V.~Aulchenko}\affiliation{Budker Institute of Nuclear Physics SB RAS, Novosibirsk 630090}\affiliation{Novosibirsk State University, Novosibirsk 630090} 
  \author{T.~Aushev}\affiliation{Moscow Institute of Physics and Technology, Moscow Region 141700} 
  \author{R.~Ayad}\affiliation{Department of Physics, Faculty of Science, University of Tabuk, Tabuk 71451} 
  \author{V.~Babu}\affiliation{Tata Institute of Fundamental Research, Mumbai 400005} 
  \author{I.~Badhrees}\affiliation{Department of Physics, Faculty of Science, University of Tabuk, Tabuk 71451}\affiliation{King Abdulaziz City for Science and Technology, Riyadh 11442} 
  \author{V.~Bansal}\affiliation{Pacific Northwest National Laboratory, Richland, Washington 99352} 
  \author{P.~Behera}\affiliation{Indian Institute of Technology Madras, Chennai 600036} 
% ...
\collaboration{The Belle Collaboration}

\begin{abstract}
A search for the decays $\decaykp$ and $\decayks$ is performed.
Evidence for the decay $\decaykp$ is found; its
significance is $4.8\sigma$. No evidence is found for $\decayks$.
The branching fraction for $\decaykp$ is measured to be
$(3.7 \err{1.0}{0.9} \err{0.8}{0.8}) \times 10^{-5}$; the upper limit for
the $\decayks$ branching fraction is $1.4 \times 10^{-5}$ at $90\%$ C.L.
In addition, a study of the $\hccha{1}$ invariant mass
distribution in the channel $\Bp \to (\hccha{1}) \kp$ results in the first
observation of the decay $\eta_c(2S) \to \hccha{1}$ with
$12.1\sigma$ significance.
The analysis is based on the 711 $\invfb$ data sample collected by the Belle
detector at the asymmetric-energy $\elp \elm$ collider KEKB at the
$\Upsilon(4S)$ resonance.
\end{abstract}

\maketitle

\section{Introduction}

The decays $\Bp \to \chi_{c0} \kp$, $\Bp \to \chi_{c2} \kp$ and
$\decaykp$ are suppressed by
factorization~\cite{Bauer:1986bm,Suzuki:2002sq}.
The decays $\Bp \to \chi_{cJ} \kp$ have been observed; the current
world-average branching fractions are
$\br(\Bp \to \chi_{c0} \kp) = (1.49\err{0.15}{0.14}) \times 10^{-4}$ and
$\br(\Bp \to \chi_{c2} \kp) = (1.1 \pm 0.4)\times10^{-5}$~\cite{Tanabashi:2018oca}.
While $\br(\Bp \to \chi_{c0} \kp)$ is smaller than the
branching fraction of the factorization-allowed process
$\br(\Bp \to \chi_{c1} \kp) = (4.84 \pm 0.23) \times 10^{-4}$, it is not
strongly suppressed. Before the first experimental searches,
this resulted in an assumption that the process $\decaykp$ may also have a
large branching fraction
$\br(\decaykp) \approx \br(\Bp \to \chi_{c0} \kp)$~\cite{Suzuki:2002sq}.

However, the decay $\decaykp$ has not been observed experimentally yet.
Neither the Belle~\cite{Fang:2006bz} nor
BaBar~\cite{Aubert:2008kp} Collaborations have
found a statistically significant signal of $\decaykp$ using the
decay mode $h_c \to \eta_c \gamma$. The current branching-fraction upper limit
of $\br(\decaykp) < 3.8 \times 10^{-5}$ at $90\%$ confidence
level (C. L.)~\cite{Tanabashi:2018oca} was obtained in the
$h_c$ search by Belle~\cite{Fang:2006bz}.

Also, the LHCb Collaboration searched for the
process $\Bp \to h_c (\to \pr \apr) \kp$~\cite{Aaij:2013rha} and set the upper
limit on the branching fraction product
$\br(\Bp \to h_c \kp) \times \br(h_c \to \pr \apr) < 6.4 \times 10^{-8}$
(95\% C. L.). However, this measurement does not result in a
stronger restriction on  $\br(\decaykp)$,
because the decay $h_c \to \pr \apr$ has never been
observed and the upper limit on its branching fraction is
$\br(h_c \to \pr \apr) < 1.5 \times 10^{-4}$
(90\% C. L.)~\cite{Tanabashi:2018oca}. Note that a newer LHCb analysis
of the same channel performed in Ref.~\cite{Aaij:2016kxn} does not update
the upper limit on $\br(\Bp \to h_c \kp) \times \br(h_c \to \pr \apr)$.

Several new theoretical predictions of $\br(\decaykp)$ were made after the
experimental upper limit was set.
The branching fraction has been calculated in
the QCD factorization approach to be $2.7 \times 10^{-5}$~\cite{Meng:2006mi}.
A calculation using perturbative QCD was performed in Ref.~\cite{Li:2006vj};
the result is $\br(\decaykp) = 3.6 \times 10^{-5}$.
Another calculation performed in Ref.~\cite{Beneke:2008pi} results in
$\br(\decaykp)$ [$\br(B^0 \to h_c K^0)$] in the interval from
$3.1 \times 10^{-5}$ to $5.7 \times 10^{-5}$
[from $2.9 \times 10^{-5}$ to $5.3 \times 10^{-5}$],
depending on the assumed value of the $c$ quark mass.
All the results mentioned above are close to each other,
and the theoretical values
of $\br(\decaykp)$ are slightly below the current experimental upper limit.
This motivates an updated study of the decays $\decaykp$, which may be able to
find them.

Here we present such an updated search for the decays $\decaykp$, and
also include a search for the decays $\decayks$.
The analysis is performed using the $711\ \invfb$ data sample collected
by the Belle detector at the asymmetric-energy $\elp \elm$ collider
KEKB~\cite{kekb}. The data sample was collected at the $\Upsilon(4S)$
resonance and contains $772 \times 10^6$ $B\bar{B}$ pairs.
The integrated luminosity is 2.8 times greater than
the luminosity used in the previous analysis~\cite{Fang:2006bz}.
For further improvement of the sensitivity,
the new analysis uses ten $\eta_c$ decay channels to reconstruct
the decay $\hcdec{0}$; only two channels were used in the old one. The
new $h_c$ decay channel $\hcdec{1}$ observed recently
by the BESIII Collaboration~\cite{Ablikim:2018ewr} is also used for its
reconstruction; in addition,
we study the decays of other charmonium states to $\hccha{1}$.
The discrimination of the signal and background events is improved
by performing a multivariate analysis.

\section{The Belle Detector}

The Belle detector is a large-solid-angle magnetic
spectrometer that consists of a silicon vertex detector (SVD),
a 50-layer central drift chamber (CDC), an array of
aerogel threshold Cherenkov counters (ACC),
a barrel-like arrangement of time-of-flight
scintillation counters (TOF), and an electromagnetic calorimeter (ECL)
comprised of CsI(Tl) crystals located inside
a superconducting solenoid coil that provides a 1.5~T
magnetic field.  An iron flux-return located outside of
the coil is instrumented to detect $K_L^0$ mesons and to identify
muons.  The detector
is described in detail elsewhere~\cite{Belle}.
Two inner detector configurations were used. A 2.0 cm radius beampipe
and a 3-layer silicon vertex detector were used for the first sample
of 140 $\invfb$, while a 1.5 cm radius beampipe, a 4-layer
silicon detector and a small-cell inner drift chamber were used to record
the remaining data~\cite{svd2}.

We use a {\sc geant}-based Monte Carlo (MC) simulation~\cite{geant} to model
the response of the detector, identify potential backgrounds and
determine the acceptance. The MC simulation includes run-dependent
detector performance variations and background conditions.  Signal MC
events are generated with {\sc EvtGen}~\cite{evtgen}
in proportion to the relative luminosities of the
different running periods.

\section{Event selection}

We select events of the type $\decaykp$ and $\decayks$.
Inclusion of charge-conjugate modes is implied hereinafter.
The reconstruction is performed with a conversion from Belle to Belle II
data format~\cite{Gelb:2018agf}.

All tracks are required to originate from the interaction point region:
we require $dr < 0.2\ \cm$ and $|dz| < 2\ \cm$, where
$dr$ and $dz$ are the cylindrical coordinates of the point of the
closest approach of the track to the beam axis
(the $z$ axis of the laboratory reference frame coincides with
the positron-beam axis).

Charged $\pi$, $K$ mesons and protons are identified using
likelihood ratios
$R_{h_1/h_2} = \mathcal{L}_{h_1}/(\mathcal{L}_{h_1}+\mathcal{L}_{h_2})$,
where $h_1$ and $h_2$ are the particle-identification hypotheses ($\pi$, $K$,
or $p$) and $\mathcal{L}_{h_i}$ are their corresponding likelihoods.
The likelihoods are calculated from the combined
time-of-flight information from the TOF, the number of photoelectrons from
the ACC and $dE/dx$ measurements in the CDC.
We require $R_{K/\pi} > 0.6$ for $K$ candidates, $R_{\pi/K} > 0.6$ for
$\pi$ candidates, and $R_{p/\pi} > 0.6$, $R_{p/K} > 0.6$ for $p$ candidates.
The identification efficiency of the above requirements varies in the ranges
(94 -- 99)\%, (84 -- 93)\%, and (90 -- 98)\%
 for $\pi$, $K$, and $p$, respectively,
depending on the $h_c$ or $\eta_c$ decay channel.
The misidentification probability
for the background particles that are not $\pi$, $K$, and $p$,
varies in the ranges (25 -- 49)\%, (4.9 -- 11.3)\%, and (0.5 -- 1.9)\%,
respectively. Without the electron background, which is rejected
as described below, the $\pi$ fake rate drops to (20 -- 35)\%.

Electron candidates are identified as CDC charged tracks
that are matched to electromagnetic showers in the ECL.
The track and ECL cluster matching quality,
the ratio of the electromagnetic shower energy to the track momentum,
the transverse shape of the shower, the ACC light yield, and the
track $dE/dx$ ionization are used in our electron identification
criteria. A similar likelihood ratio is constructed:
$R_e = \mathcal{L}_e / (\mathcal{L}_e + \mathcal{L}_h)$,
where $\mathcal{L}_e$ and $\mathcal{L}_h$ are the likelihoods for electrons and
charged hadrons ($\pi$, $K$ and $p$), respectively~\cite{Hanagaki:2001fz}.
An electron veto ($R_e < 0.9$) is imposed on $\pi$, $K$, and $p$ candidates.
It is not applied for the $\ks$ and $\Lambda$ daughter tracks, because
they have independent selection criteria. For the $h_c$ or $\eta_c$ decay
channels other than $\etacdec{2}$ and $\etacdec{16}$, the electron veto
rejects from 3.5 to 15\% of the background events, while its signal
efficiency is not less than 97.5\%.

Photons are identified as ECL electromagnetic showers that
have no associated charged tracks detected in the CDC. The
shower shape is required to be consistent with that of a photon.

The $\piz$ candidates are reconstructed via their
decay to two photons. The photon energies in the laboratory frame are
required to be greater than $30\ \mev$.
The $\piz$ invariant mass is required to satisfy
$|M_{\piz} - m_{\piz}| < 15\ \mevcc$. Here and elsewhere,
$M_\text{particle}$ denotes the reconstructed invariant mass of the specified
particle and $m_\text{particle}$ stands for the nominal mass of this
particle~\cite{Tanabashi:2018oca}.
This requirement corresponds approximately
to a $3\sigma$ mass window around the nominal mass.

The $V^0$-particle ($\ks$ and $\Lambda$) candidates 
are reconstructed from pairs of oppositely charged tracks that are
assumed to be $\pip \pim$ and $p \pim$ for $\ks$ and $\Lambda$, respectively.
We require $|M_{\ks} - m_{\ks}| < 20\ \mevcc$ and
$|M_{\Lambda} - m_{\Lambda}| < 10\ \mevcc$, corresponding
approximately to $5.5\sigma$ mass windows in both cases.
The $V^0$ candidates are selected by a neural network
using the following input
variables: the $V^0$ candidate momentum, decay angle,
flight distance in the $xy$ plane, the angle between the $V^0$
momentum and the direction from the interaction point to the $V^0$ vertex,
the shortest $z$ distance between the two daughter tracks,
their radial impact parameters, and numbers of hits in the SVD and CDC.
The separation of the $\ks$ and $\Lambda$ candidates is performed by another
neural network. The input variables of this network are the
momenta and polar angles of the daughter tracks in the laboratory frame,
their likelihood ratios $R_{\pi/p}$ and the $V^0$ candidate invariant mass for
the $\Lambda$ hypothesis.

The $\eta$ candidates are reconstructed in $\gamma\gamma$ and $\pip\pim\piz$
channels. The reconstructed $\eta$ candidates are denoted by the
$\eta$ decay channel as $\eta_{2\gamma}$ and $\eta_{3\pi}$. The $\eta$
invariant mass is required to satisfy
$|M_{\eta_{2\gamma}} - m_{\eta}| < 30\ \mevcc$ and
$|M_{\eta_{3\pi}} - m_{\eta}| < 15\ \mevcc$; these requirements correspond to
$2.5\sigma$ and $4\sigma$ mass windows,
respectively.

The $\eta'$ candidates are reconstructed in the $\eta \pip \pim$ decay mode.
The invariant mass window is $|M_{\eta'} - m_{\eta'}| < 15\ \mevcc$,
corresponding to a $4\sigma$ mass window.

The $\eta_c$ candidates are reconstructed in ten decay channels:
$\etaccha{0}$, $\etaccha{1}$, $\etaccha{2}$, $\kp \km \eta$,
$\etaccha{5}$, $\eta' (\to \eta \pip \pim) \pip \pim$,
$\etaccha{13}$, $\etaccha{14}$, $\etaccha{15}$, and $\etaccha{16}$.
The selected $\eta_c$ candidates
are required to satisfy $|M_{\eta_c} - m_{\eta_c}| < 50\ \mevcc$;
the mass-window width is about $1.6$ widths of the $\eta_c$.

The $h_c$ candidates are reconstructed in the $\hcdec{0}$ and $\hcdec{1}$
decay channels.
The invariant mass of the $h_c$ candidates is not restricted
for the channel $\hccha{0}$; for the channel $\hccha{1}$, it is required
to be greater than $2.7\ \gevcc$. The lower mass limit is selected to be
very low to study other charmonium states decaying to the same final state.

The $B$-meson candidates are reconstructed via the decay modes
$\decaykp$ and $\decayks$.
The $B$ candidates are selected by
their energy and the beam-energy-constrained mass. The difference of the
$B$-meson and beam energies is defined as $\de = \sum_i E_i - E_{\text{beam}}$,
where $E_i$ are the energies of the $B$ decay products in the center-of-mass
frame and $E_{\text{beam}}$ is the beam energy in the same frame.
The beam-energy-constrained mass is defined as
$\mbc = \sqrt{E_{\text{beam}}^2-(\sum_i\vec{p}_i)^2}$,
where $\vec{p}_i$ are the momenta of the $B$ decay products in the
center-of-mass frame. We retain $B$ candidates satisfying the conditions
$5.2 < \mbc < 5.3\ \gevcc$ and $|\de| < 0.2\ \gev$. A mass-constrained fit is
applied to the selected $B$-meson candidates.

In addition, for the channel $\hcdec{0}$, the
$h_c$ daughter $\gamma$ energy is required to be greater than $200\ \mev$ in
the $B$ rest frame. This requirement removes the background from low-energy
photons, including the peaking backgrounds from $B$ decays to the
same final state without the photon. The signal efficiency of this requirement
is 100\%, because the $\eta_c \gamma$ invariant mass of
all excluded events is smaller than the $h_c$ mass.

Also, the ratio of the Fox-Wolfram moments~\cite{Fox:1978vu} $F_2/F_0$ is
required to be less than 0.3.
This requirement reduces the continuum background, rejecting
from 18\% to 53\% of background events, depending on the $h_c$ or $\eta_c$
decay channel. Its signal efficiency is from 93.3\% to 96.3\%.

\section{Multivariate analysis and optimization of the selection requirements}

\subsection{General analysis strategy and data samples}

To improve the separation of the signal and background events, we perform a
multivariate analysis followed by an optimization of selection requirements.
The first stages of the analysis are performed individually for $\hcdec{1}$ and
each $\eta_c$ decay channel for the $h_c$ candidates reconstructed in the
$\hccha{0}$ mode (the channels $\eta_c \to \kp \km \eta$ and
$\eta_c \to \eta' (\to \eta \pip \pim) \pip \pim$
are optimized separately for $\eta_{2\gamma}$ and $\eta_{3\pi}$).
They include the determination of two-dimensional
$(\de,\mbc)$ resolution and the distribution of the background
in $(\de,\mbc)$, and the multivariate-analysis stage.
The optimization of the selection requirements uses the results of all
initial stages as its input. The resolution is used to determine
the expected number of the signal events and the distribution of
the background in $(\de,\mbc)$ is used to determine the expected number
of the background events in the signal region. The optimization is performed
individually for the channel $\hcdec{1}$ and globally for all
$\eta_c$ decay channels for the channel $\hcdec{0}$.
The data selected using the resulting channel-dependent
criteria are merged into a single sample for the $\hcdec{0}$ channel.
The final fit is performed simultaneously to the $\hcdec{0}$
and $\hcdec{1}$ samples.

The experimental data are used for determination of the $(\de,\mbc)$
distribution, selection of the background samples for the neural network, and
final fit to the selected events.
During the development of the analysis procedure, the $h_c$ signal region
was excluded to avoid bias of the $h_c$ significance.
The final fit described in Sec.~\ref{sec:fit} was performed on MC
pseudoexperiments generated in accordance with the fit result without the $h_c$
mixed with the $\decaykp$ ($\decayks$) signal MC.
The $h_c$ signal region is defined by
\begin{equation}
\sqrt{\left(\frac{\de}{\sigma_{\de}}\right)^2 +
\left(\frac{\mbc - m_B}{\sigma_{\mbc}}\right)^2} < 3,
\label{eq:blinded_region_de_mbc}
\end{equation}
where $\sigma_{\de} = 18\ \mev$ and $\sigma_{\mbc} = 2.5\ \mevcc$ are
the approximate resolutions in $\de$ and $\mbc$, respectively, and
\begin{equation}
\begin{aligned}
3.50 < M_{h_c} < 3.55\ \gevcc & \text{ for $\hcdec{0}$,} \\
3.515 < M_{h_c} < 3.535\ \gevcc & \text{ for $\hcdec{1}$.} \\
\end{aligned}
\label{eq:blinded_region_hc}
\end{equation}
After completion of the analysis procedure development, this requirement
is no longer used.

The signal MC is used for the determination of the resolution and
the selection of the signal samples for the neural network.
The signal MC is generated using the known information about the angular or
invariant-mass distributions of the decay products if it is possible;
otherwise, uniform distributions are assumed.
The angular distribution is known for the channel $\hcdec{0}$.
It does not have any free parameters and is proportional to
$\sin^2 \theta_{h_c}$, where $\theta_{h_c}$ is the $h_c$ helicity angle
that is defined as the angle between $-\vec{p}_{B}$ and $\vec{p}_{\eta_c}$,
where $\vec{p}_{B}$ and $\vec{p}_{\eta_c}$ are the momenta of the $B$
and $\eta_c$ in the $h_c$ rest frame, respectively.
In addition, the $\eta_c$ decay resonant structure is taken
into account if it is known. The distributions for the channels
$\etaccha{0}$, $\etaccha{1}$, $\etaccha{2}$, $\etaccha{3}$, and $\etaccha{4}$
are based on the results of a Dalitz plot analysis performed
in Ref.~\cite{Lees:2014iua}. The contributions of intermediate $\phi$
resonances are taken into account for the channel $\etaccha{5}$ based on
the world-average branching fractions from Ref.~\cite{Tanabashi:2018oca}.

\subsection{Resolution}

\begin{figure}
\includegraphics[width=8.0cm]{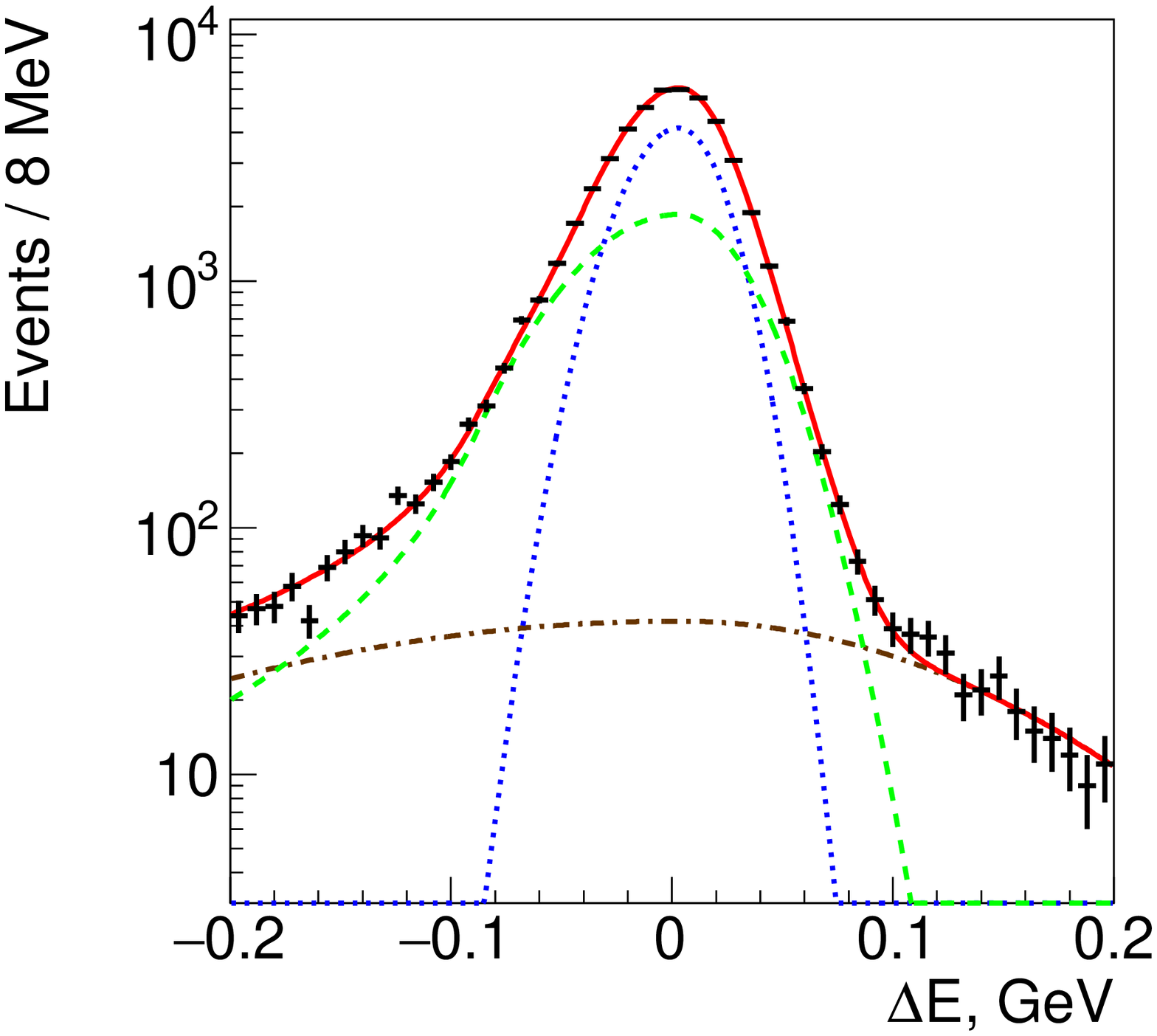}
\includegraphics[width=8.0cm]{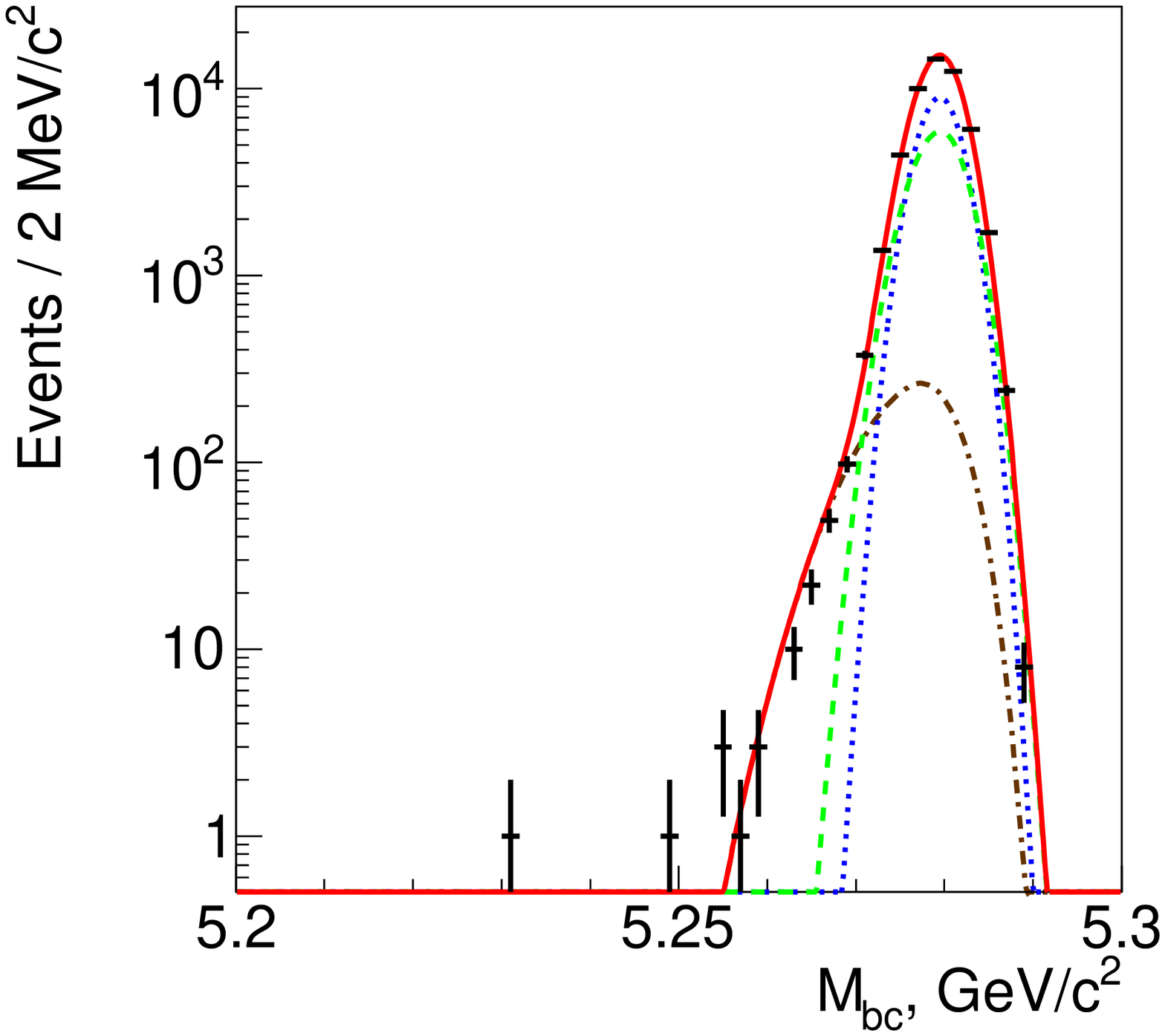}
\caption{Projections of the resolution fit results onto $\de$ and
$\mbc$ for the channel $\decaykp$ with $\optdec{1}$.
The red solid line is the fit result, the green
dashed line is the Crystal Ball component, the blue dotted line is the first
Gaussian component, and the brown dash-dotted line is the second Gaussian
component.}
\label{fig:resolution}
\end{figure}

The resolution is parameterized by the function
\begin{equation}
\begin{aligned}
S(\de, \mbc) = & N_{\text{CB}} F_{\text{CB}}(x_1) G_a^{(1 2)}(y_1) \\
& + N_\text{G1} G_a^{(2 1)}(x_2) G_a^{(2 2)}(y_2) \\
& + N_\text{G2} G_a^{(3 1)}(x_3) G_a^{(3 2)}(y_3), \\
\label{eq:signal_pdf}
\end{aligned}
\end{equation}
where $F_{\text{CB}}$ is an asymmetric Crystal Ball
function~\cite{skwarnicki},
$G_a^{(i j)}$ are asymmetric Gaussian functions,
$N_{\text{CB}}$, $N_\text{G1}$ and $N_\text{G2}$ are normalizations and
$x_i$ and $y_i$ ($i$ = 1, 2, 3) are rotated variables
that are given by
\begin{equation}
\begin{pmatrix}
x_i \\
y_i \\
\end{pmatrix}
=
\begin{pmatrix}
\cos\alpha_i & \sin\alpha_i \\
-\sin\alpha_i & \cos\alpha_i \\
\end{pmatrix}
\begin{pmatrix}
\de - (\de)_0 \\
\mbc - (\mbc)_0 \\
\end{pmatrix}.
\end{equation}
Here, ($(\de)_0$, $(\mbc)_0$) is the central point and
$\alpha_i$ is the rotation angle. The central point is
the same for all three components.
The resolution is determined from a binned maximum likelihood fit
to signal MC events. Example resolution fit results (for the channel $\decaykp$
with $\optdec{1}$) are shown in Fig.~\ref{fig:resolution}.

\subsection{Fit to the $(\de,\mbc)$ distribution}

The $(\de,\mbc)$ distribution is fitted in order to estimate the expected
number of the background events in the signal region.
The distribution is fitted to the function
\begin{equation}
N_S S(\de,\mbc) + B(\de,\mbc),
\end{equation}
where $N_S$ is the number of signal events and $B$ is the background density
function that is given by
\begin{equation}
\begin{aligned}
B(\de,\mbc) = & \sqrt{m_0 - \mbc} \exp[-a (m_0 - \mbc)] \\
& \times P_3(\de,\mbc), \\
\end{aligned}
\label{eq:background_density}
\end{equation}
where $m_0$ is the threshold mass,
$a$ is a rate parameter and $P_3$ is a two-dimensional third-order polynomial.
The region with $\de < -0.12\ \gev$ is excluded for the
channel $\hcdec{1}$ because of the presence of peaking backgrounds from
partially reconstructed B decays with an additional $\pi$ meson.

\begin{figure}
\includegraphics[width=8.0cm]{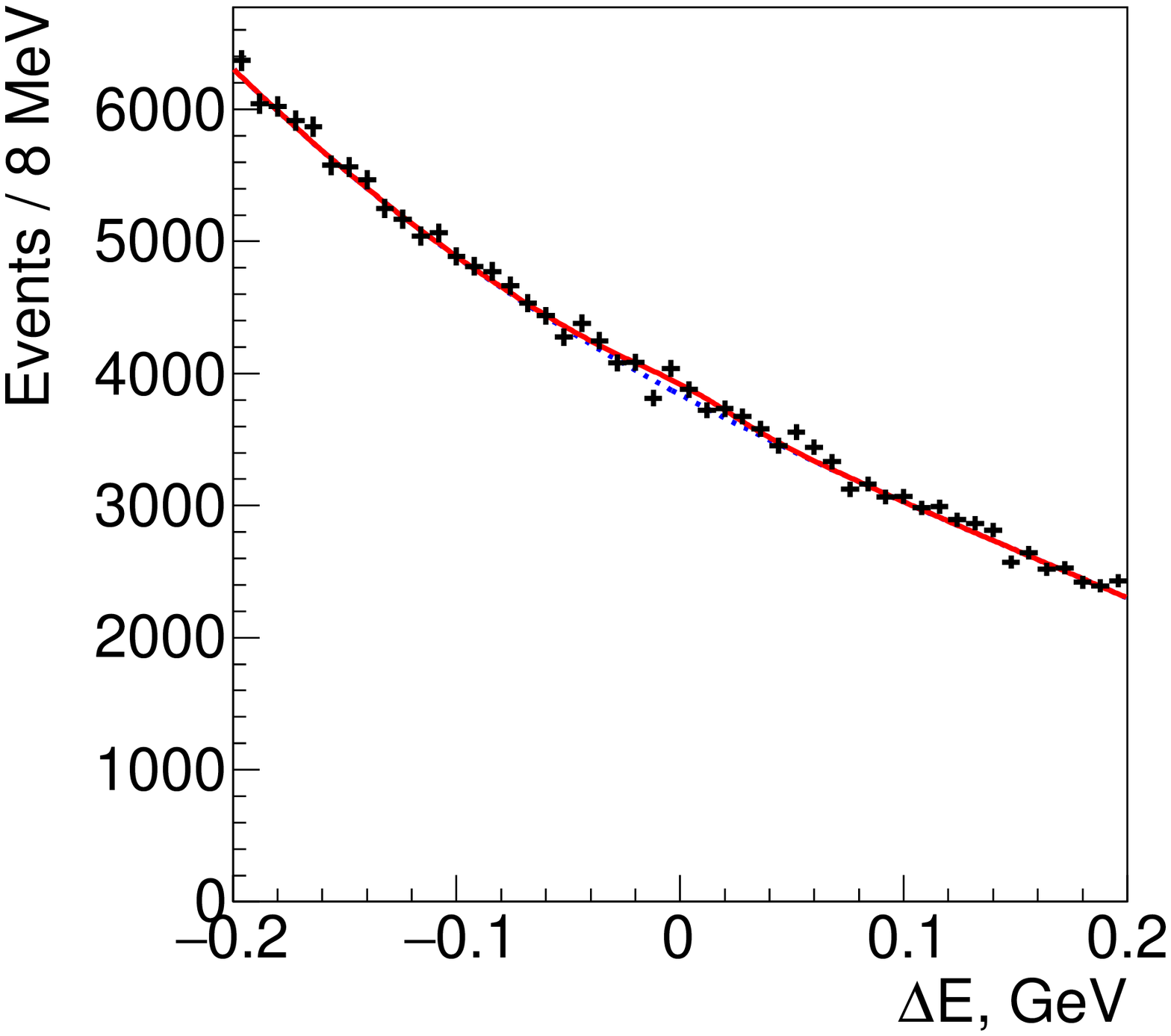}
\includegraphics[width=8.0cm]{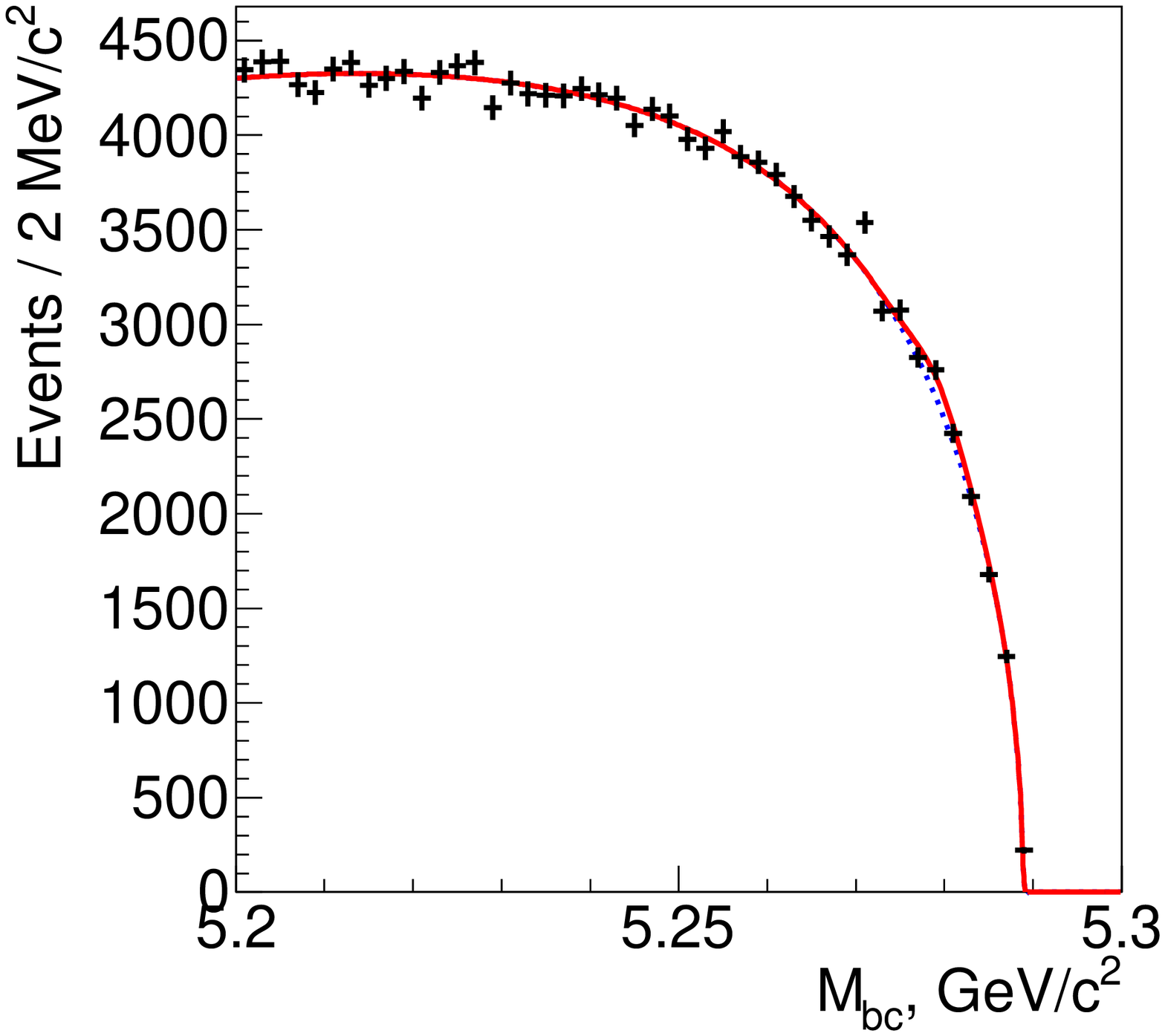}
\caption{Projections of the results of the fit to 
the $(\de,\mbc)$ distribution onto $\de$
(with $\mbc > 5.272\ \gevcc$) and $\mbc$ (with $|\de| < 20\ \mev$)
for the channel $\decaykp$ with $\optdec{1}$. The red solid line is the fit
result, and the blue dotted line is the background. Since there is no
significant signal before the optimization of the selection requirements and
for the entire $\hccha{0}$ mass range, the two lines almost coincide.}
\label{fig:background}
\end{figure}

Example $(\de,\mbc)$ fit results (for the channel $\decaykp$ with $\optdec{1}$)
are shown in Fig.~\ref{fig:background}.

\subsection{Multivariate analysis}
\label{sec:mva}

To improve the separation of signal and background events,
we perform a multivariate analysis for each individual channel.
The algorithm used for the multivariate analysis is the multilayer
perceptron (MLP) neural network implemented in the {\sc tmva}
library~\cite{tmva}.
The following variables are always included in the neural network:
the angle between the thrust axes of the $B$ candidate and the remaining
particles in the event, the angle between the thrust axes of all tracks
and all photons in the event, the ratio of the Fox-Wolfram moments
$F_2/F_0$, the $B$ production angle, and the vertex fit quality.
For the $h_c$ candidates reconstructed in the $\hccha{0}$ channel,
the MLP also includes the $h_c$ helicity angle, the $\eta_c$ mass, and
the number of $\piz$ candidates that include the $h_c$ daughter photon
as one of their daughters (separately for two groups of $\piz$ candidates
with the energy of another photon less and greater than $100\ \mev$).

For the channels $\etacdec{0}$, $\etacdec{1}$, and $\etacdec{2}$, two
invariant masses of the $\eta_c$ daughter particle pairs
(both $K \pi$ combinations) are added to the neural network.

The following particle identification variables are included into
the neural network if there are corresponding charged particles
in the final state: the minimum likelihood ratio $R_{K/\pi}$ of
the $h_c$ daughter kaons, the minimum of the two likelihood ratios
$R_{p/K}$, $R_{p/\pi}$ of the $h_c$ daughter protons, and
$R_{K/\pi}$ for the $B$ daughter $\kp$ (for the channel $\decaykp$).
Here, the $h_c$ daughters may be either direct (from the decay $\hcdec{1}$)
or indirect (the $\eta_c$ daughters for the $h_c$ candidates reconstructed in
the $\hccha{0}$ mode).

If there is a $\piz$ or $\eta$ decaying to $\gamma \gamma$ in the final state,
four additional variables are added: the $\piz$ ($\eta$) mass,
the minimal energy of the $\piz$ ($\eta$) daughter photons in the
laboratory frame, and the number of $\piz$ candidates that include a
$\piz$ ($\eta$) daughter photon as one of their daughters (for each of
the $\piz$ ($\eta$) daughter photons).
If there is an $\eta$ reconstructed in the $\pip\pim\piz$ decay mode, then
only its mass is added to the MLP.
If the $\eta_c$ has a daughter $\eta'$, then the mass of the $\eta'$
candidate is also included to the neural network.

The training and testing signal samples are taken from
the signal MC.
The background sample is taken from a two-dimensional $(\de,\mbc)$ sideband.
For the channel $\decaykp$, the sideband is defined as
\begin{equation}
3 < \sqrt{\left(\frac{\de}{\sigma_{\de}}\right)^2 +
\left(\frac{\mbc - m_B}{\sigma_{\mbc}}\right)^2} < 8.
\end{equation}
The background sample is divided into
training and testing samples of equal size.

The channel $\decayks$ has a small number of background events. In order
to avoid overtraining, the background region for this channel is redefined.
It includes all selected events except the central region defined by
Eq.~\eqref{eq:blinded_region_de_mbc}.
In addition, the MLP internal architecture is changed. Instead of the default
{\sc tmva} neural network with two hidden layers,
only one hidden layer is used.

\begin{figure}
\includegraphics[width=8.0cm]{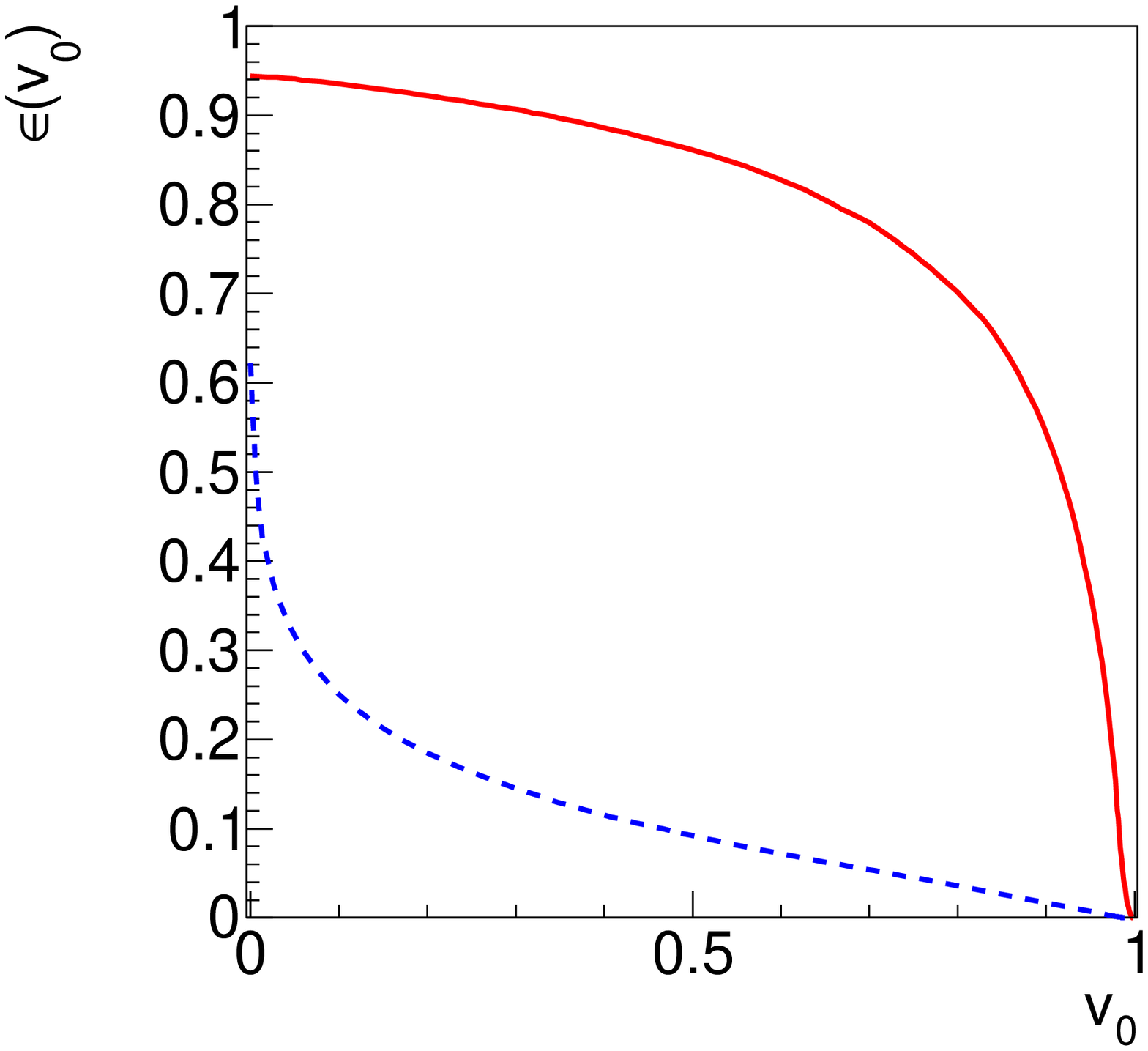}
\caption{Efficiency of the MLP output requirement ($v > v_0$)
for the channel $\decaykp$ with $\optdec{1}$.
The red solid line is the signal efficiency and the blue dashed line
is the background efficiency.}
\label{fig:mva_output}
\end{figure}

The resulting efficiency of the requirement ($v > v_0$) on the MLP output
variable $v$ for the training sample is shown in
Fig.~\ref{fig:mva_output} for the channel $\decaykp$ with $\optdec{1}$.
Note that the efficiency is given by
\begin{equation}
\epsilon(v_0) = \epsilon_\text{MLP}(v_0) \times \epsilon_\text{multiple}(v_0),
\end{equation}
where $\epsilon$ is the full efficiency, $\epsilon_\text{MLP}$ is the raw MLP
output requirement efficiency and $\epsilon_\text{multiple}$ is the
best-candidate selection efficiency. Because of the correction
by $\epsilon_\text{multiple}$, the efficiency at the minimal MLP output value
$v_\text{min}$ is not 1 but rather $\epsilon_\text{multiple}(v_\text{min})$.

The best-candidate selection is performed for each of the
multivariate-analysis channels separately in the following way.
The selected $(\de,\mbc)$ region is subdivided into three bins in both $\de$
and $\mbc$. The selection is performed for each of the bins separately.
The candidate with the largest MLP output is selected.
One of the bins ($-67 < \de < 67\ \mev$, $5.267 < \mbc < 5.3\ \gevcc$)
always contains the entire signal region selected by the optimization
procedure as described in Sec.~\ref{sec:global_optimization}.
Thus, the signal region of the final data sample does not contain multiple
candidates that originate from the same $h_c$ channel. However, multiple
candidates from different $h_c$ channels are possible.

The best-candidate selection efficiency increases for larger
values of the MLP output cutoff value $v_0$. For the $v_0$ values obtained as
the result of the optimization of the selection requirements as described in
Sec.~\ref{sec:global_optimization}, the selection procedure removes from $3$
to 15\% of data events, depending on the multivariate-analysis channel.

\subsection{Optimization of the selection requirements}
\label{sec:global_optimization}

Optimization of the selection requirements is performed by maximizing the value
\begin{equation}
F_\text{opt} = \frac{\sum\limits_i N_\text{sig}^{(i)}}
{\displaystyle{\frac{a}{2}} + \sqrt{\sum\limits_i N_\text{bg}^{(i)}}},
\label{eq:optfun}
\end{equation}
where $i$ is the channel index, $N_\text{sig}^{(i)}$
is the expected number of the signal events for the $i$-th channel,
$N_\text{bg}^{(i)}$ is the expected number of the background events
in the signal region,
and $a=3$ is the target significance.
This optimization method is based on Ref.~\cite{Punzi:2003bu}.

The signal region is defined as
\begin{equation}
\left(\frac{\de}{R_{\de}^{(i)}}\right)^2 +
\left(\frac{\mbc-m_B}{R_{\mbc}^{(i)}}\right)^2 < 1.
\end{equation}
where $R_{\de}^{(i)}$ and $R_{\mbc}^{(i)}$
are the half-axes of the signal region ellipse.
The parameters determined by the optimization are
$R_{\de}^{(i)}$, $R_{\mbc}^{(i)}$,
and the minimal value of the MLP output ($v_0^{(i)}$)
for each channel.

The expected number of signal events for $\decaykp$ is calculated as
\begin{equation}
\begin{aligned}
N_\text{sig}^{(i)} = & 2 N_{\Upsilon(4S)} \br(\Upsilon(4S)\to\Bp\Bm)
\br(\decaykp) \\
& \times
\br(h_c \to i) \epsilon_\text{SR}^{(i)}
\epsilon_S^{(i)}(v_0^{(i)}), \\
\end{aligned}
\end{equation}
where $N_{\Upsilon(4S)}$ is the number of $\Upsilon(4S)$ events,
$\br(h_c \to i)$ is the branching fraction of the $h_c$ to its $i$-th decay
channel,
$\epsilon_\text{SR}^{(i)}$ is the reconstruction efficiency for the
specific signal region $\text{SR}$, and
$\epsilon_S^{(i)}(v_0^{(i)})$ is the efficiency of the requirement ($v > v_0$)
on the MLP output variable $v$ for the signal events.
The number of $\Upsilon(4S)$ events is assumed to be equal to the number of
$B \bar{B}$ pairs; the branching fraction $\br(\Upsilon(4S)\to\Bp\Bm)$
is calculated under the same assumption~\cite{Tanabashi:2018oca}.
The signal-region-dependent reconstruction efficiency is calculated as
\begin{equation}
\epsilon_\text{SR}^{(i)} = \epsilon_R^{(i)}
\int\limits_\text{SR} S_i(\de,\mbc) d \de d \mbc,
\end{equation}
where $\epsilon_R^{(i)}$ is the reconstruction efficiency,
and $S_i$ is the signal PDF for $i$-th $h_c$
decay channel (the integral of $S_i$ over the signal region is the
efficiency of the signal region selection).
The unknown branching fraction
$\br(\decaykp)$ can be set to an arbitrary value because the maximum of
$F_\text{opt}$ does not depend on it. The expected number of signal events
for $\decayks$ is calculated similarly.

The expected number of background events is calculated as
\begin{equation}
N_\text{bg}^{(i)} = \epsilon_B^{(i)}(v_0^{(i)})
\frac{N_{h_c\text{ region}}}{N_\text{full}}
\int\limits_\text{SR} B_i(\de,\mbc) d \de d\mbc,
\end{equation}
where $\epsilon_B^{(i)}(v_0^{(i)})$ is the efficiency of the MLP output
requirement for the background events,
$N_{h_c\text{ region}}$ is the number of
background events in the $h_c$ region defined by
Eq.~\eqref{eq:blinded_region_hc}, $N_\text{full}$ is the full number
of the background events, and
$B_i$ is the background density function defined in
Eq.~\eqref{eq:background_density} for $i$-th $h_c$ decay channel.

The optimization is performed separately for two channel groups.
The first group includes the multivariate-analysis channels corresponding to
the decay $\hcdec{0}$; the index $i$ runs over all $\eta_c$ decay channels.
The second group consists of the single channel $\hcdec{1}$. The separate
optimization is required by the difference of further data processing:
the data from the first group of channels are
combined into a single $\hcdec{0}$ data sample, while the $\hcdec{1}$ data
are fitted with another function, as described below in Sec.~\ref{sec:fit}.
The optimization results are shown in Table~\ref{tab:optimization}.
We also check the improvement achieved by MLP usage
by changing the selection method to rectangular cuts.
The values of $F_\text{opt}$ are found to be about 30\% and 10\% smaller
for the channels $\hcdec{0}$ and $\hcdec{1}$, respectively.

\begin{table*}
\caption{Results of the optimization of the selection requirements.
The signal-region half-axes $R_{\de}^{(i)}$ ($R_{\mbc}^{(i)}$)
are in $\mev$ ($\mevcc$); all other values are dimensionless.}
\begin{tabular}{c|c|c|c|c|c|c|c|c|c|c|c|c}
\hline\hline
\multirow{3}{*}{Channel} & \multicolumn{6}{|c|}{$\decaykp$} & \multicolumn{6}{|c}{$\decayks$} \\
\cline{2-13}
 & \multicolumn{3}{|c|}{Parameters} & \multicolumn{3}{|c}{Efficiency}
 & \multicolumn{3}{|c|}{Parameters} & \multicolumn{3}{|c}{Efficiency} \\
\cline{2-13}
 & $R_{\de}^{(i)}$ & $R_{\mbc}^{(i)}$ & $v_0^{(i)}$ & $\epsilon_\text{SR}^{(i)}$ & $\epsilon_S^{(i)}(v_0^{(i)})$ & $\epsilon_B^{(i)}(v_0^{(i)})$
 & $R_{\de}^{(i)}$ & $R_{\mbc}^{(i)}$ & $v_0^{(i)}$ & $\epsilon_\text{SR}^{(i)}$ & $\epsilon_S^{(i)}(v_0^{(i)})$ & $\epsilon_B^{(i)}(v_0^{(i)})$ \\
\hline
\multicolumn{13}{c}{Channel group 1: $\hcdec{0}$} \\
\hline
$\optcha{0}$ & 32.7 & 4.82 & 0.804 & 6.27\% & 59.5\% & 5.08\%
             & 34.2 & 4.97 & 0.702 & 4.31\% & 69.8\% & 8.55\% \\
$\optcha{1}$ & 36.2 & 3.90 & 0.958 & 4.27\% & 31.7\% & 0.56\%
             & 43.5 & 4.54 & 0.942 & 3.38\% & 39.3\% & 0.85\% \\
$\optcha{2}$ & 42.3 & 4.49 & 0.976 & 1.79\% & 17.8\% & 0.18\%
             & 35.9 & 4.07 & 0.954 & 1.05\% & 35.9\% & 0.64\% \\
$\optcha{3}$ & 34.4 & 4.16 & 0.977 & 4.21\% & 20.2\% & 0.22\%
             & 37.8 & 4.49 & 0.967 & 3.10\% & 28.0\% & 0.41\% \\
$\optcha{4}$ & 24.9 & 3.59 & 0.978 & 1.75\% & 29.7\% & 0.23\%
             & 33.2 & 4.55 & 0.986 & 1.50\% & 23.7\% & 0.17\% \\
$\optcha{5}$ & 25.3 & 4.13 & 0.770 & 4.89\% & 53.2\% & 6.69\%
             & 29.9 & 4.80 & 0.734 & 3.71\% & 56.9\% & 8.49\% \\
$\optcha{9}$ & 30.5 & 4.21 & 0.958 & 2.69\% & 40.5\% & 0.63\%
             &  32.0 & 4.50 & 0.946 & 1.87\% & 45.6\% & 0.96\% \\
$\optcha{10}$ & 26.8 & 4.16 & 0.990 & 1.01\% & 29.2\% & 0.13\%
              & 24.6 & 3.87 & 0.986 & 0.59\% & 32.3\% & 0.26\% \\
$\optcha{13}$ & 38.9 & 5.48 & 0.654 & 17.70\% & 75.7\% & 10.66\%
              & 42.5 & 5.85 & 0.513 & 12.47\% & 82.8\% & 15.43\% \\
$\optcha{14}$ & 30.2 & 3.75 & 0.954 & 4.65\% & 30.3\% & 0.50\%
              & 31.8 & 4.01 & 0.934 & 3.33\% & 39.4\% & 0.93\% \\
$\optcha{15}$ & 24.1 & 4.03 & 0.912 & 6.31\% & 30.0\% & 1.53\%
              & 24.3 & 4.09 & 0.860 & 4.23\% & 41.8\% & 3.26\% \\
$\optcha{16}$ & 40.4 & 5.66 & 0.727 & 4.04\% & 70.6\% & 6.79\%
              & 41.5 & 5.19 & 0.586 & 2.65\% & 76.3\% & 11.24\% \\
\hline
\multicolumn{13}{c}{Channel group 2: $\hcdec{1}$} \\
\hline
$\optcha{17}$ & 13.5 & 4.36 & 0.598 & 14.81\% & 64.6\% & 18.40\%
              & 13.8 & 4.56 & 0.519 & 10.30\% & 71.2\% & 24.20\% \\
\hline\hline
\end{tabular}
\label{tab:optimization}
\end{table*}

\begin{figure}
\includegraphics[width=8.0cm]{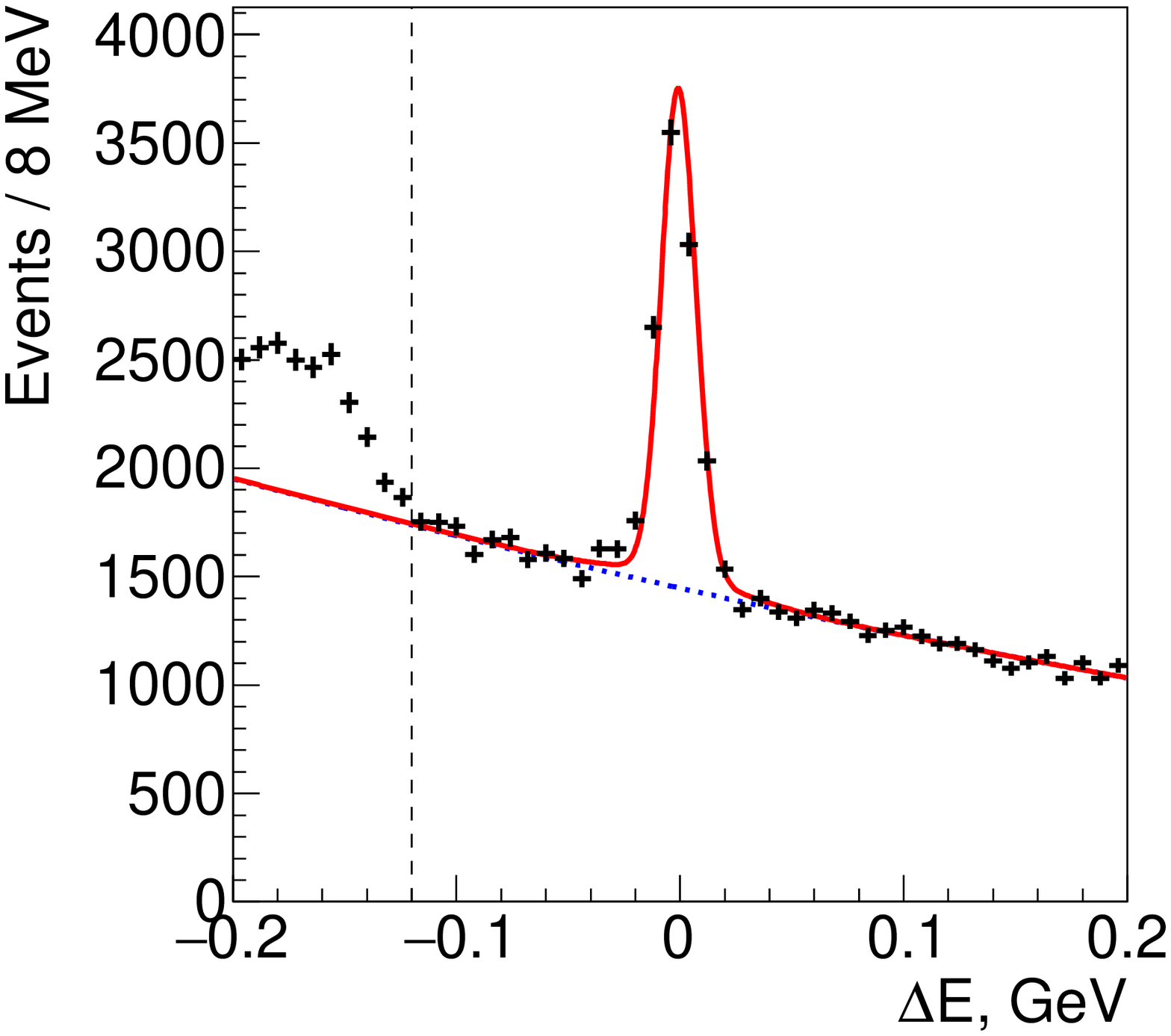}
\includegraphics[width=8.0cm]{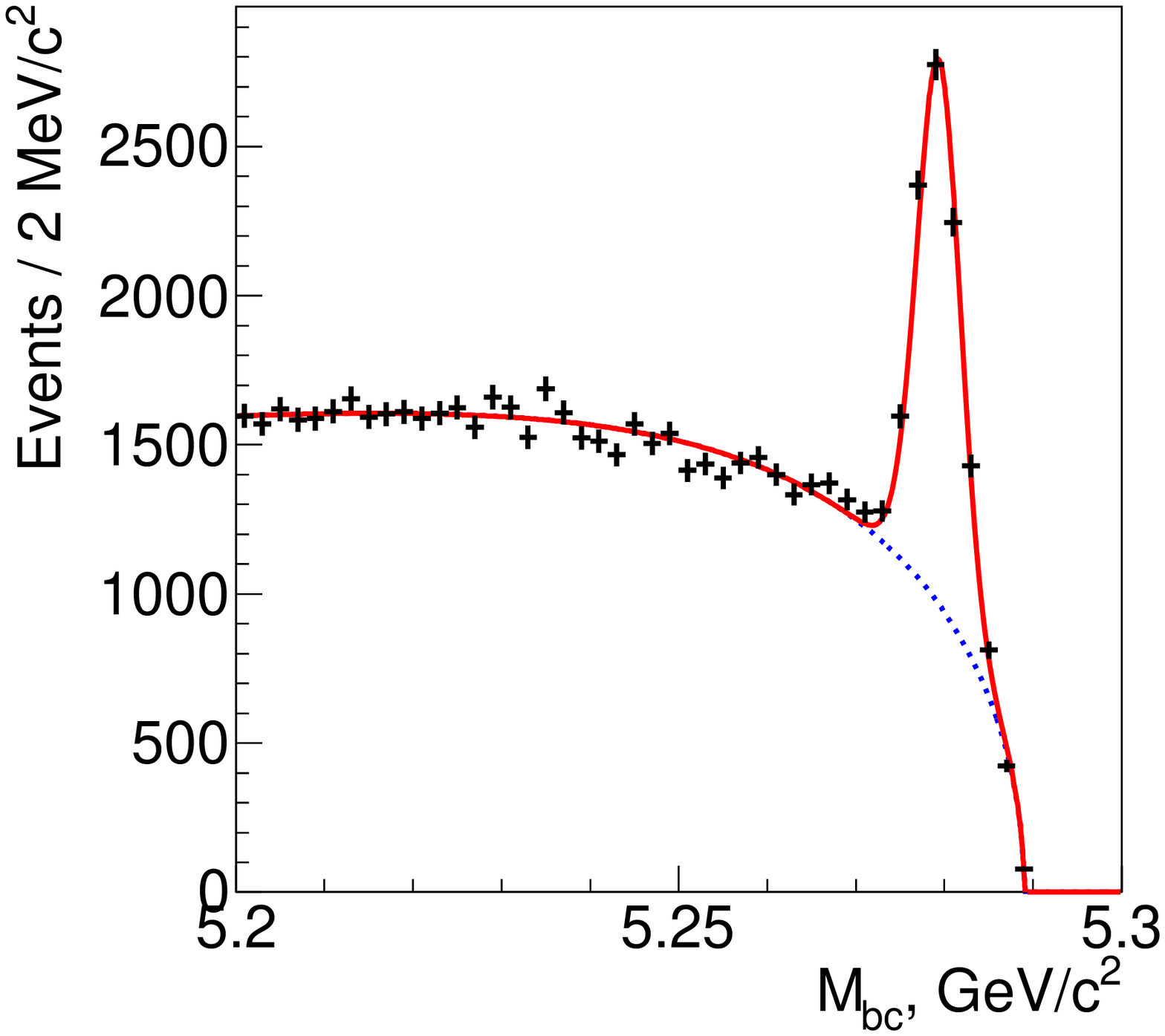}
\caption{Projections of the results of the fit to
the $(\de,\mbc)$ distribution onto $\de$
(with $\mbc > 5.272\ \gevcc$) and $\mbc$ (with $|\de| < 20\ \mev$)
for the channel $\decaykp$ with $\hcdec{1}$ after the application of the final
MLP output selection criterion. The red solid line is the fit
result, and the blue dotted line is the background.
The region with $\de < -0.12\ \gev$ is excluded
from the fit because of the presence of peaking backgrounds from
partially reconstructed B decays with an additional $\pi$ meson.
The cutoff value is marked by a vertical dashed line.
}
\label{fig:background_opt}
\end{figure}

After the optimization, the resulting selection criteria are applied. The
selected events for the channel $\hcdec{0}$ are merged.
The resolution and distribution in $(\de,\mbc)$ are determined again
for the $\hcdec{1}$ sample, since the knowledge of the background
distribution in $(\de,\mbc)$ is necessary for the final fit described in
Sec.~\ref{sec:fit}. The $(\de,\mbc)$ fit results
are shown in Fig.~\ref{fig:background_opt}.

\subsection{Resolution in $M_{h_c}$}

\begin{figure}
\includegraphics[width=8.0cm]{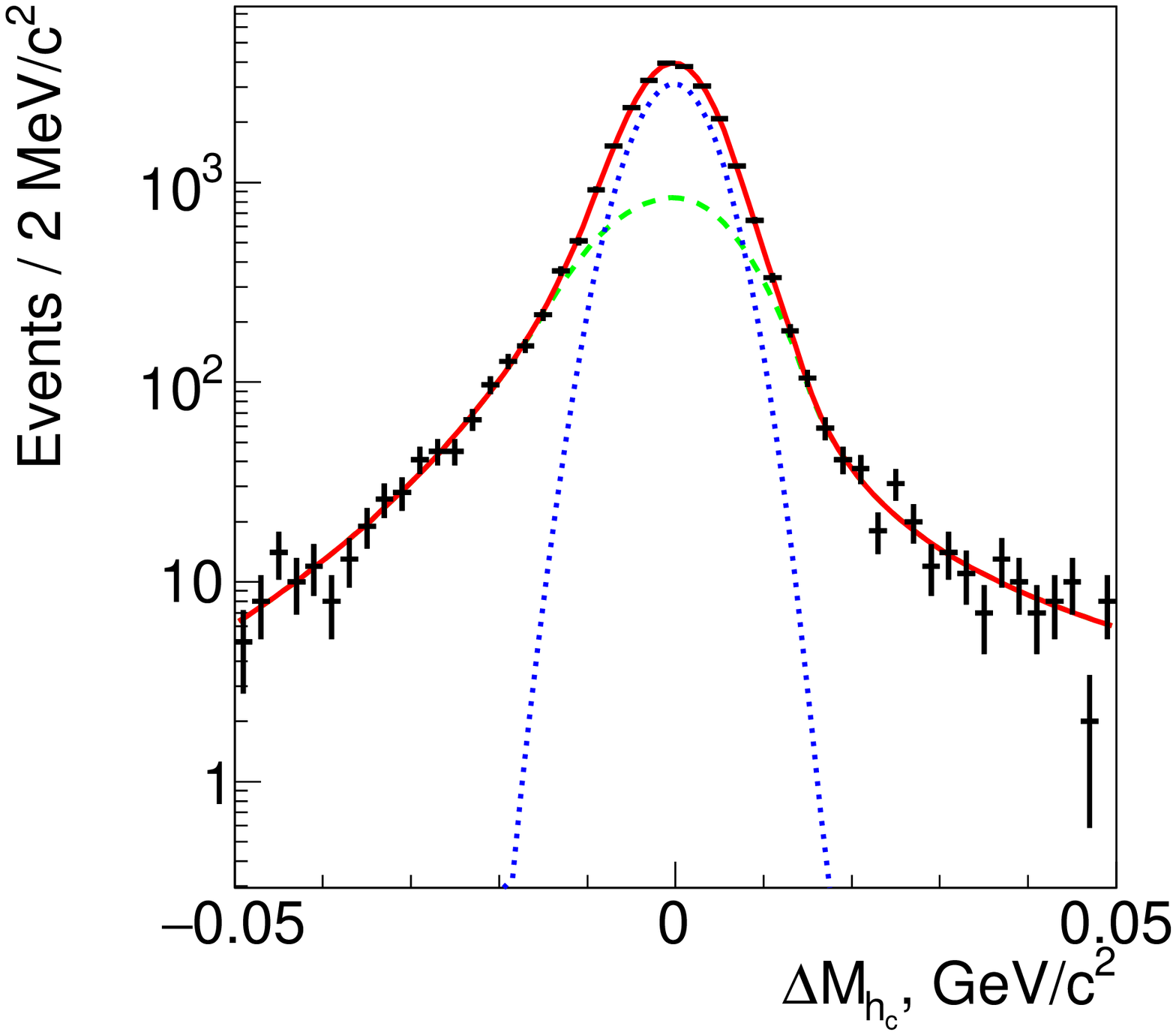}
\caption{Resolution in $M_{h_c}$ for the channel $\decaykp$ with $\hcdec{0}$.
The red solid line is the fit result,
the green dashed line is the Crystal Ball component, and the blue dotted line
is the Gaussian component.}
\label{fig:resolution_mhc}
\end{figure}

The resolution in $M_{h_c}$ is determined from a fit to
the combined $\hcdec{0}$ or $\hcdec{1}$ signal MC samples
with $\eta_c$ decaying to the reconstructed channels only. All final
selection criteria are applied.
The distribution of the difference of
the reconstructed and true masses is fitted to a sum of an
asymmetric Gaussian and asymmetric double-sided Crystal Ball functions:
\begin{equation}
R_{h_c}{(\Delta M)} = N [F_{\text{CB}}(\Delta M) f_{\text{CB}} +
G_a(\Delta M) (1 - f_{\text{CB}})],
\end{equation}
where $\Delta M$ is the difference of the reconstructed and
true $h_c$ masses, $N$ is the common normalization and
$f_{\text{CB}}$ is the Crystal Ball fraction.
Example resolution fit results (for the channel $\decaykp$ with $\hcdec{0}$)
are shown in Fig.~\ref{fig:resolution_mhc}.

\section{Fit to the data}
\label{sec:fit}

\subsection{Default model}

For the $\hcdec{0}$ final sample, the distribution in the $h_c$ mass in the
$(\de,\mbc)$ sideband cannot be used to constrain the background level
in the signal region because of the presence of peaking
backgrounds, such as the background from $B$ decays
to a similar final state with a $\piz$ instead of the $h_c$ daughter $\gamma$.
If the second photon from this $\piz$ has a small energy, then
the $\de$ and $\mbc$ values are close to 0 and the $B$ mass,
respectively. Thus, the fit is based on the signal distribution only for the
channel $\hcdec{0}$.

For the channel $\hcdec{1}$, there is a signal from $B$ decays to the same
final state ($\Bp \to \hccha{1} \kp$ or $\Bz \to \hccha{1} \ks$)
that do not proceed via any charmonium state, called the
noncharmonium signal hereinafter. Because of the possible interference of
the charmonium and noncharmonium signals, the distribution of the noncharmonium
signal in the $\hccha{1}$ invariant mass needs to be determined by the fit.
Thus, both signal and background distributions are included into the fit
for the channel $\hcdec{1}$.

We perform a simultaneous extended unbinned maximum likelihood fit to
the $\hcdec{0}$
signal, $\hcdec{1}$ background, and $\hcdec{1}$ signal distributions.
The charmonium states are represented
by the Breit-Wigner amplitude:
\begin{equation}
A_{R}(M_R) = \frac{1}{M_R^2 - m_R^2 + i M_R \Gamma_R},
\end{equation}
where $M_R$ is the invariant mass, $m_R$ is the nominal mass,
and $\Gamma_R$ is the width of the resonance $R$.
The signal-region density function for the channel $\hcdec{0}$ is given by
\begin{equation}
S_{\hccha{0}}(M) = \left(N_{h_c} |A_{h_c}(M)|^2\right) \otimes
R_{h_c}^{(\hccha{0})}(\Delta M) + P_2(M),
\end{equation}
where $N_{h_c}$ is the number of signal events, $R_{h_c}^{(\hccha{0})}$
is the $h_c$ mass resolution for the channel $\hccha{0}$,
and $P_2$ is a second-order polynomial.
The background density function $B_{\hccha{1}}(M)$
for the channel $\hcdec{1}$ is a third-order polynomial.
The signal density function for the channel $\hcdec{1}$ is given by
\begin{equation}
\begin{aligned}
&
S_{\hccha{1}}(M) = \\
& \qquad
\Big(|P_3(M) + \sum\limits_{R = \eta_c,\chi_{c0},\eta_c(2S)}
\sqrt{N_R} e^{i \varphi_R} A_R(M)|^2 \\
& \qquad
+ \sum\limits_{R = J/\psi,\chi_{c1},h_c,\chi_{c2},\psi(2S)}
N_R |A_R(M)|^2\Big) \\
& \qquad
\otimes R_{h_c}^{(\hccha{1})}(\Delta M), \\
\end{aligned}
\end{equation}
where $P_3$ is a third-order polynomial representing the noncharmonium signal.
The wide states are added coherently to the signal density function,
while the states that are narrower than the resolution are added incoherently.
The amplitudes are normalized in such a way that all the parameters $N_R$
represent the yields of the corresponding states.
The signal distribution is fitted to the function
\begin{equation}
S_{\hccha{1}}(M) + w B_{\hccha{1}}(M),
\end{equation}
where $w$ is the weight of the background events in the signal region
that is calculated as the ratio of integrals of the background distribution
in $(\de,\mbc)$ over the signal and background regions.
The model described above is the default one; additional
models are considered to study systematic uncertainties.
In the default model, the masses and widths of all resonances
are fixed to their world-average values~\cite{Tanabashi:2018oca};
all other parameters are free.

The best-candidate selection procedure described in Sec.~\ref{sec:mva}
guarantees that there are no multiple candidates in the $\hcdec{1}$
signal sample, but multiple candidates in the $\hcdec{0}$ signal sample are
possible if they originate from different $\eta_c$ decay channels. However,
the fraction of the events with multiple candidates is found to be
negligibly small. No events with multiple candidates (for the $h_c$ masses
within the default fitting regions) are observed for
both $\decaykp$ and $\decayks$ channels.

\begin{figure*}
\includegraphics[width=8.0cm]{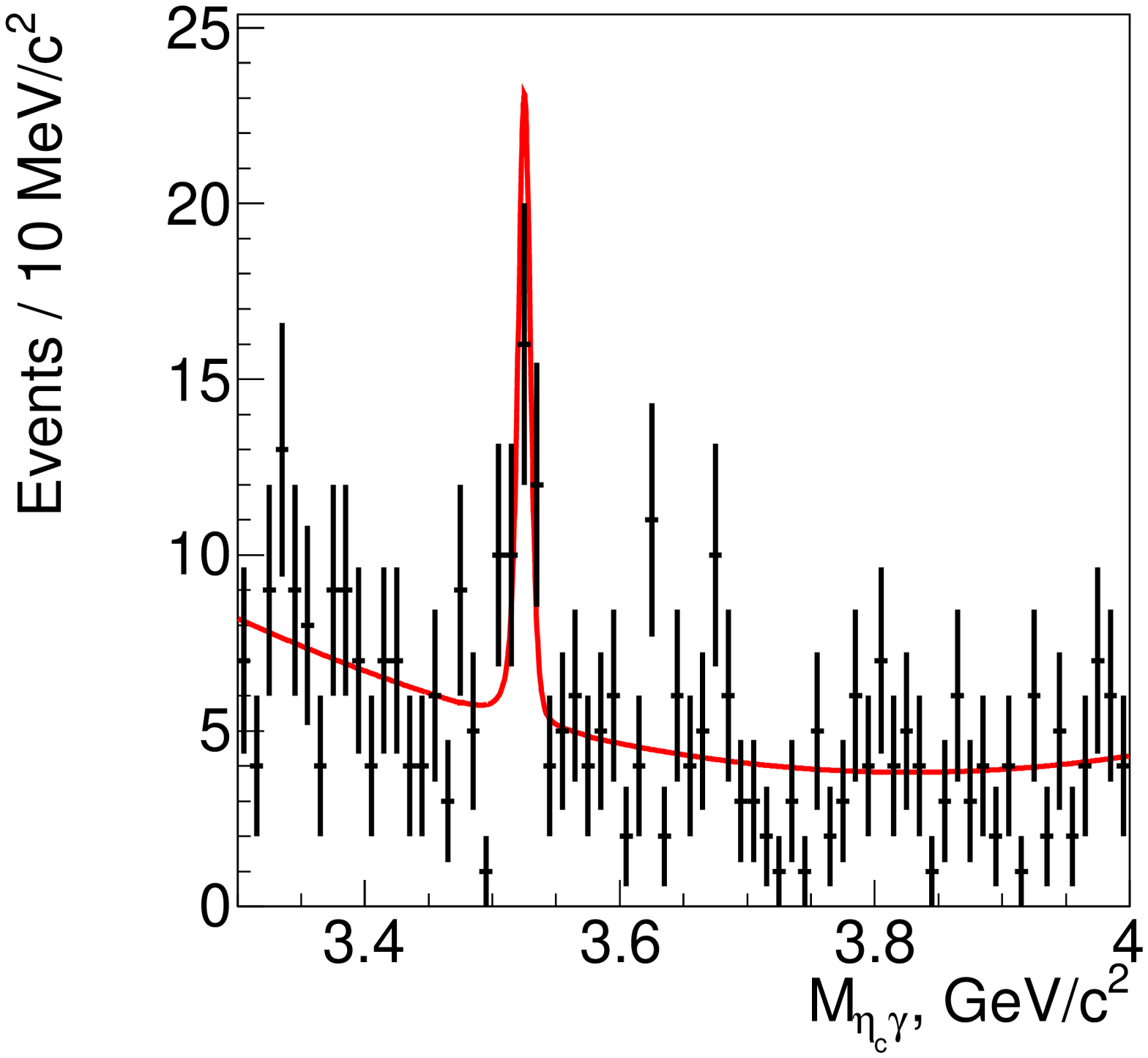}
\includegraphics[width=8.0cm]{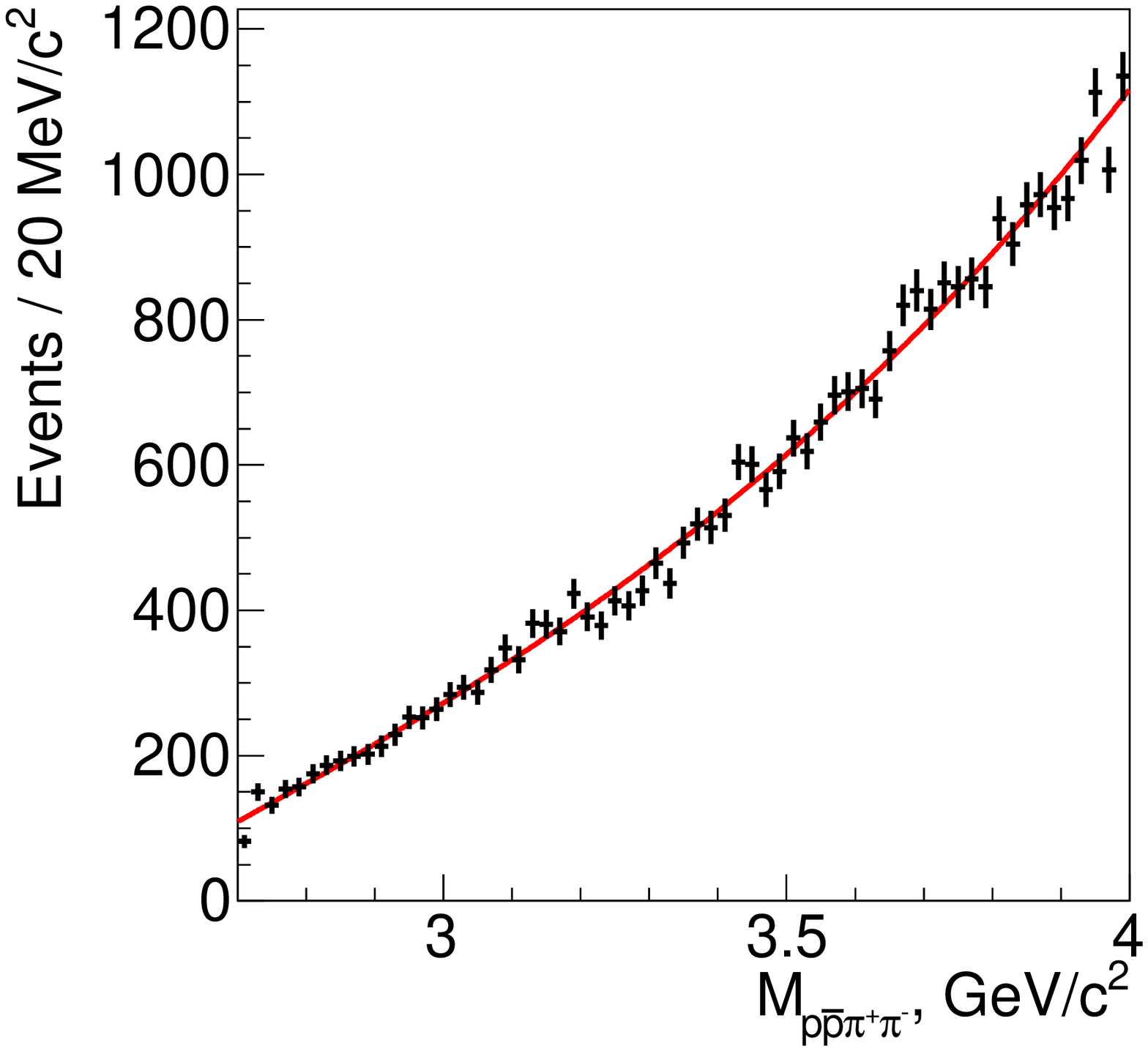} \\
\includegraphics[width=8.0cm]{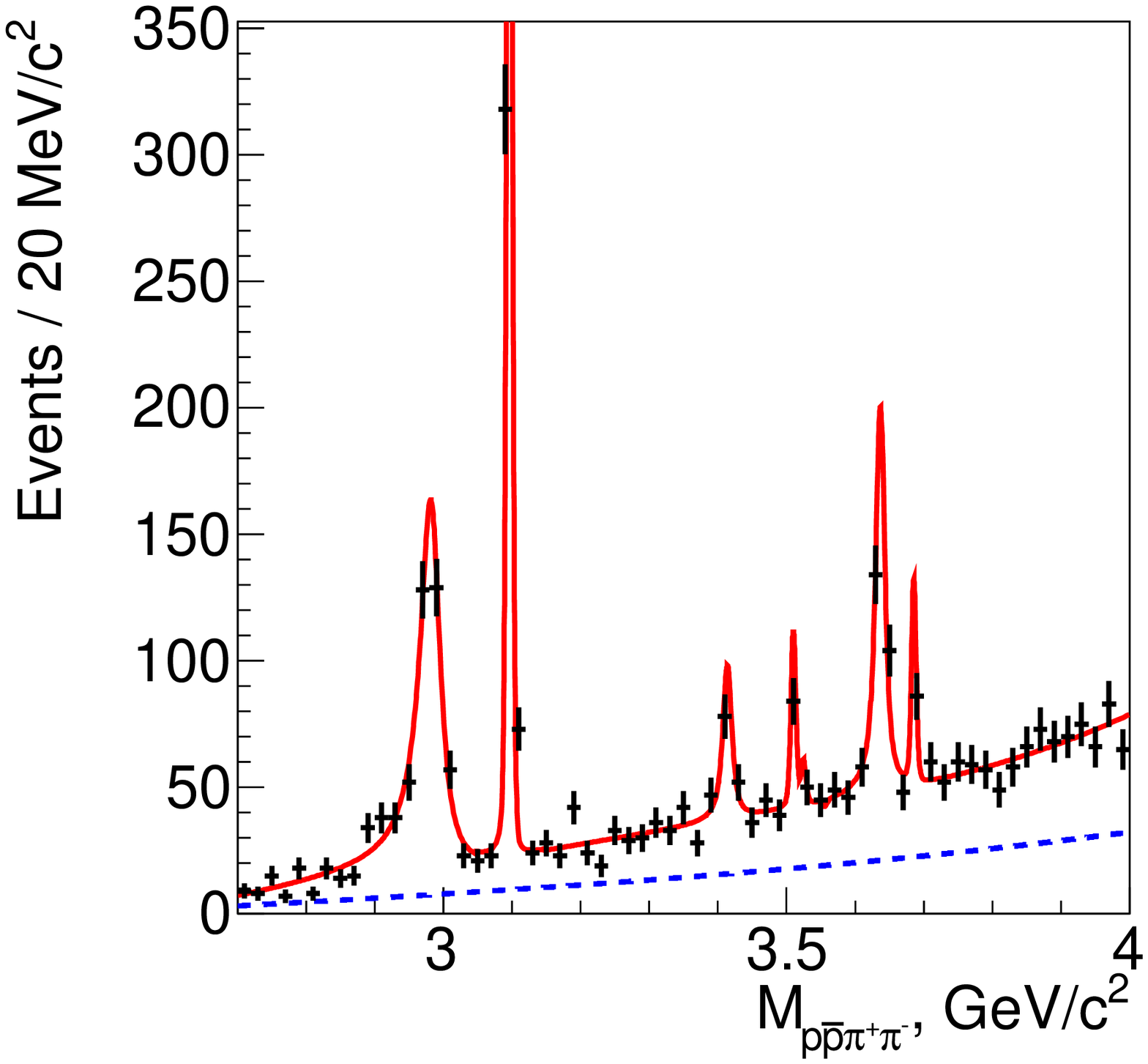}
\includegraphics[width=8.0cm]{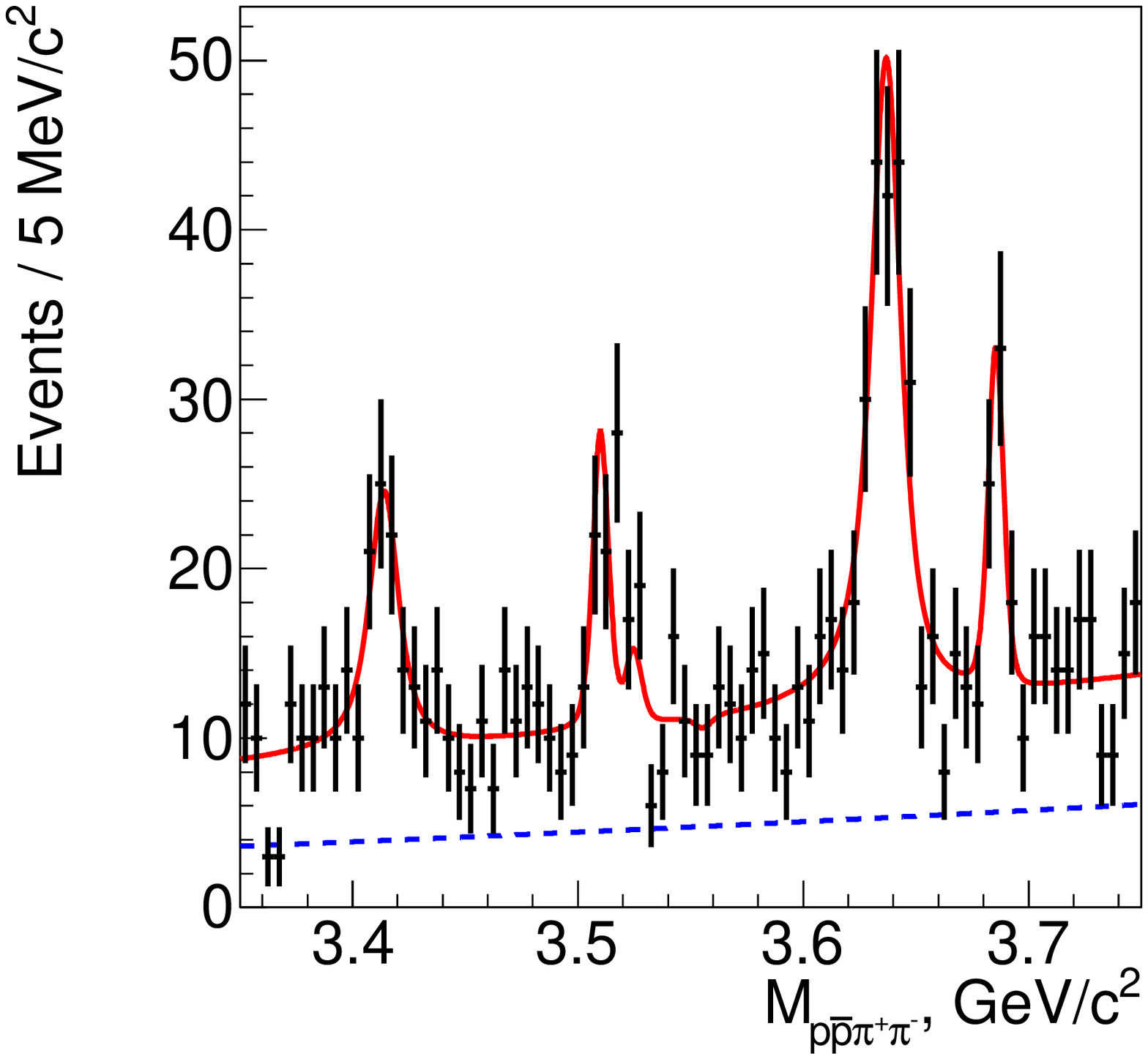} \\
\caption{Fit results in the $\decaykp$ channel:
$\hcdec{0}$ signal (top left), $\hcdec{1}$ background (top right),
$\hcdec{1}$ signal in the entire fitting region (bottom left)
and in the $\chi_{cJ}$ region (bottom right).
The red solid line is the fit result and the blue dashed line is the
background. The maximum of the $\hccha{1}$ signal fit result
at the $J/\psi$ peak is
more than two times greater than the number of data events in the corresponding
bin because the $J/\psi$ peak is narrower than the bin size. Thus,
a part of the fit result at the $J/\psi$ peak is not shown.}
\label{fig:hckp_fit_data}
\end{figure*}

\begin{figure*}
\includegraphics[width=8.0cm]{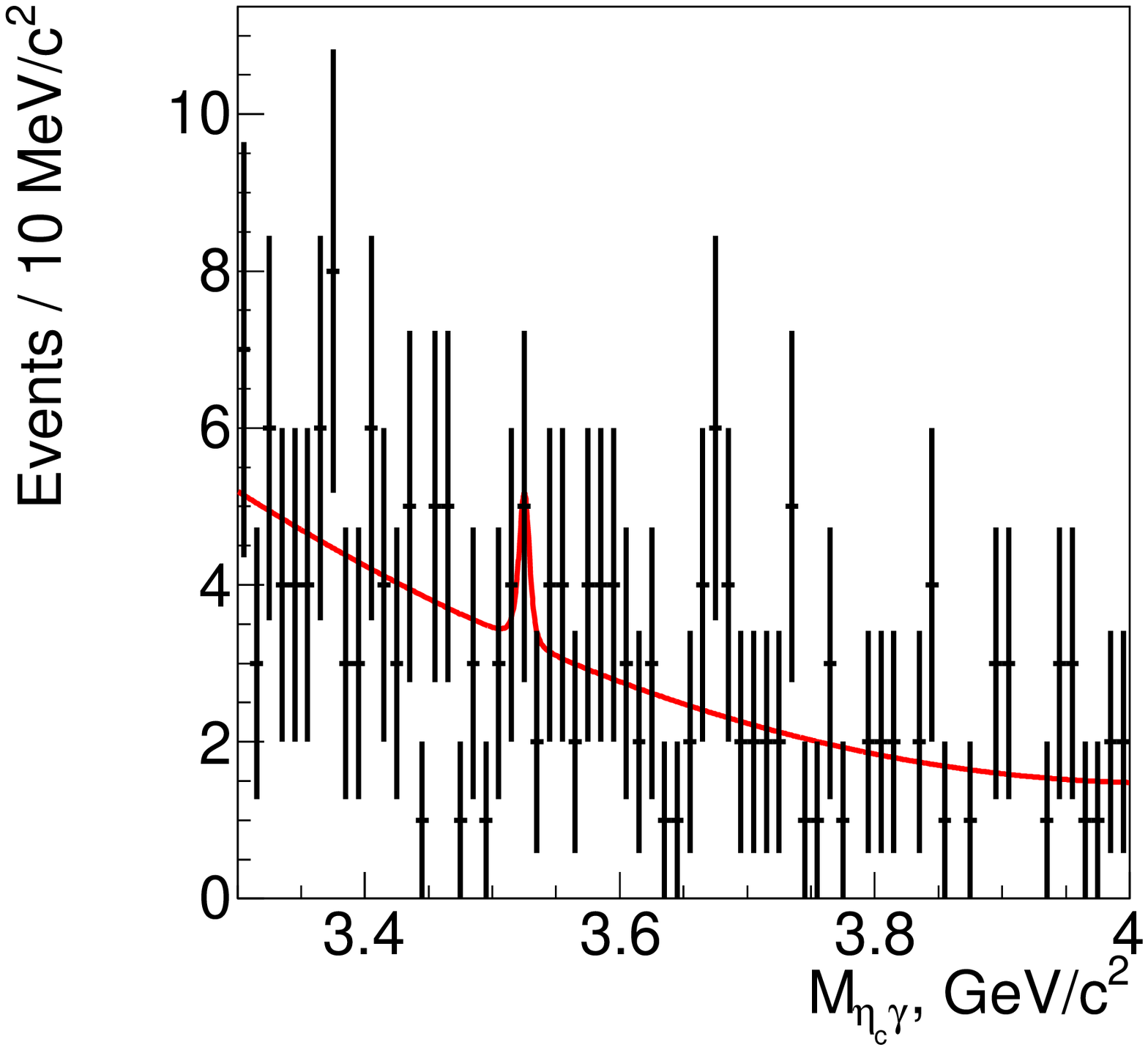}
\includegraphics[width=8.0cm]{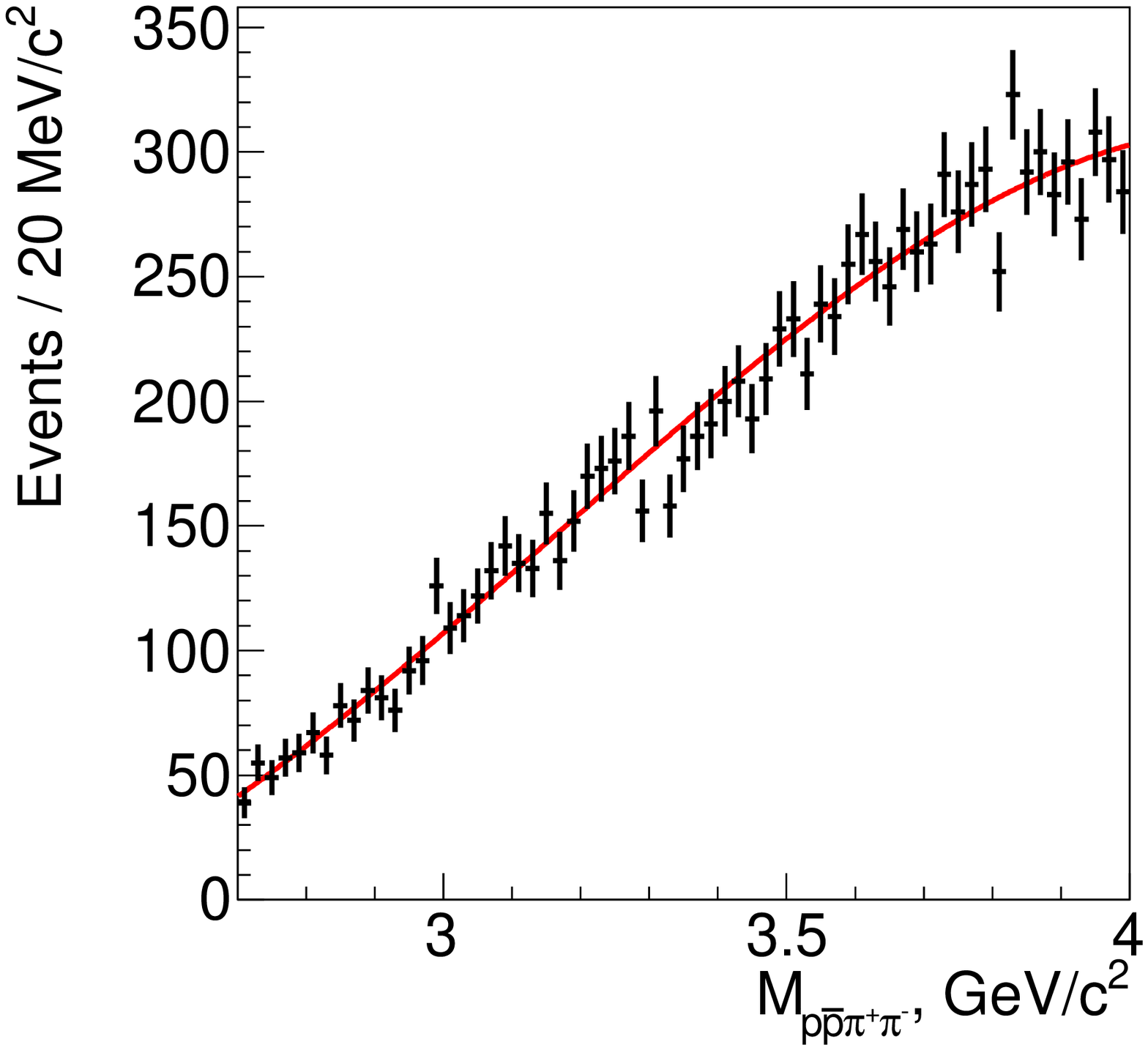} \\
\includegraphics[width=8.0cm]{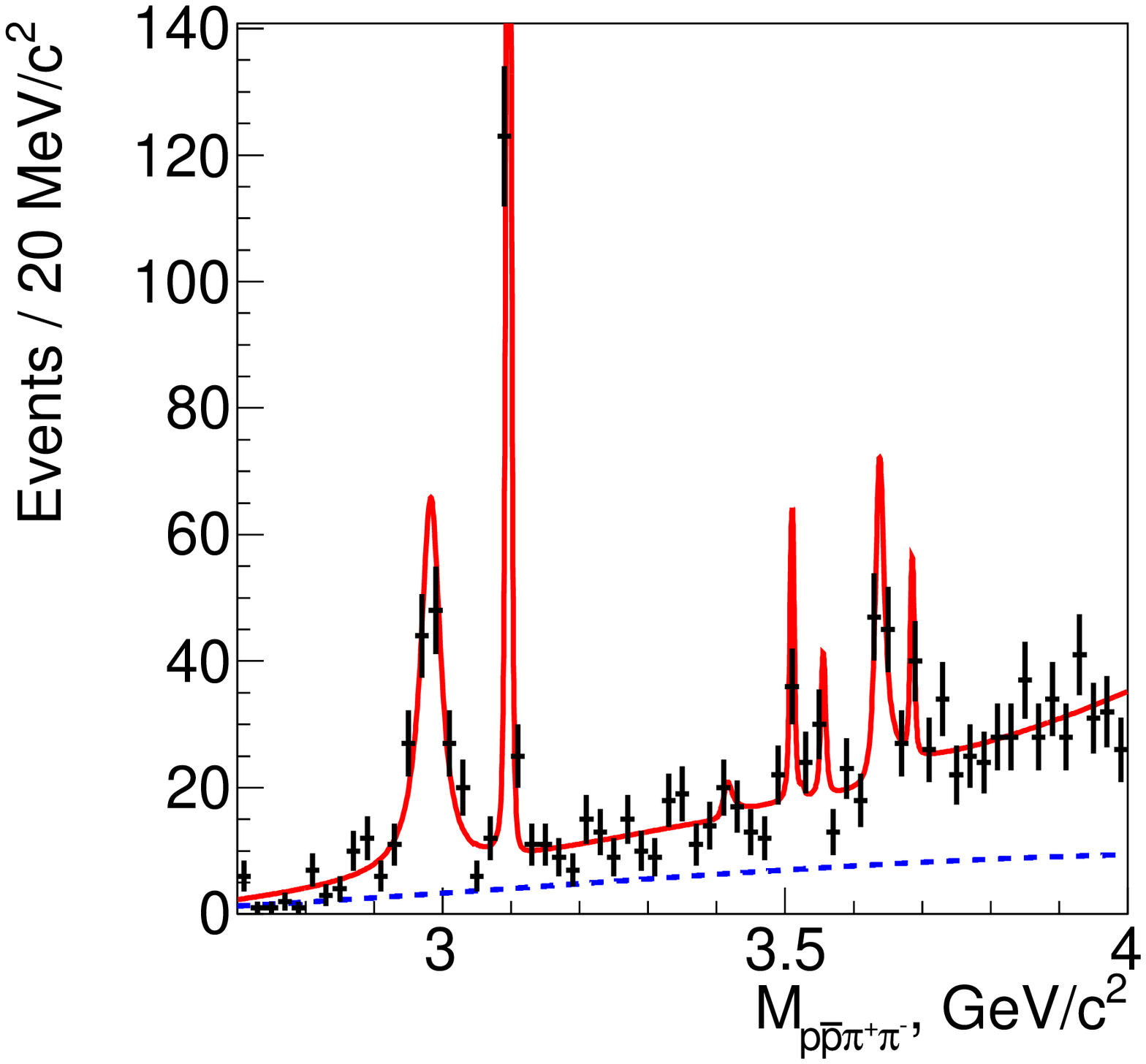}
\includegraphics[width=8.0cm]{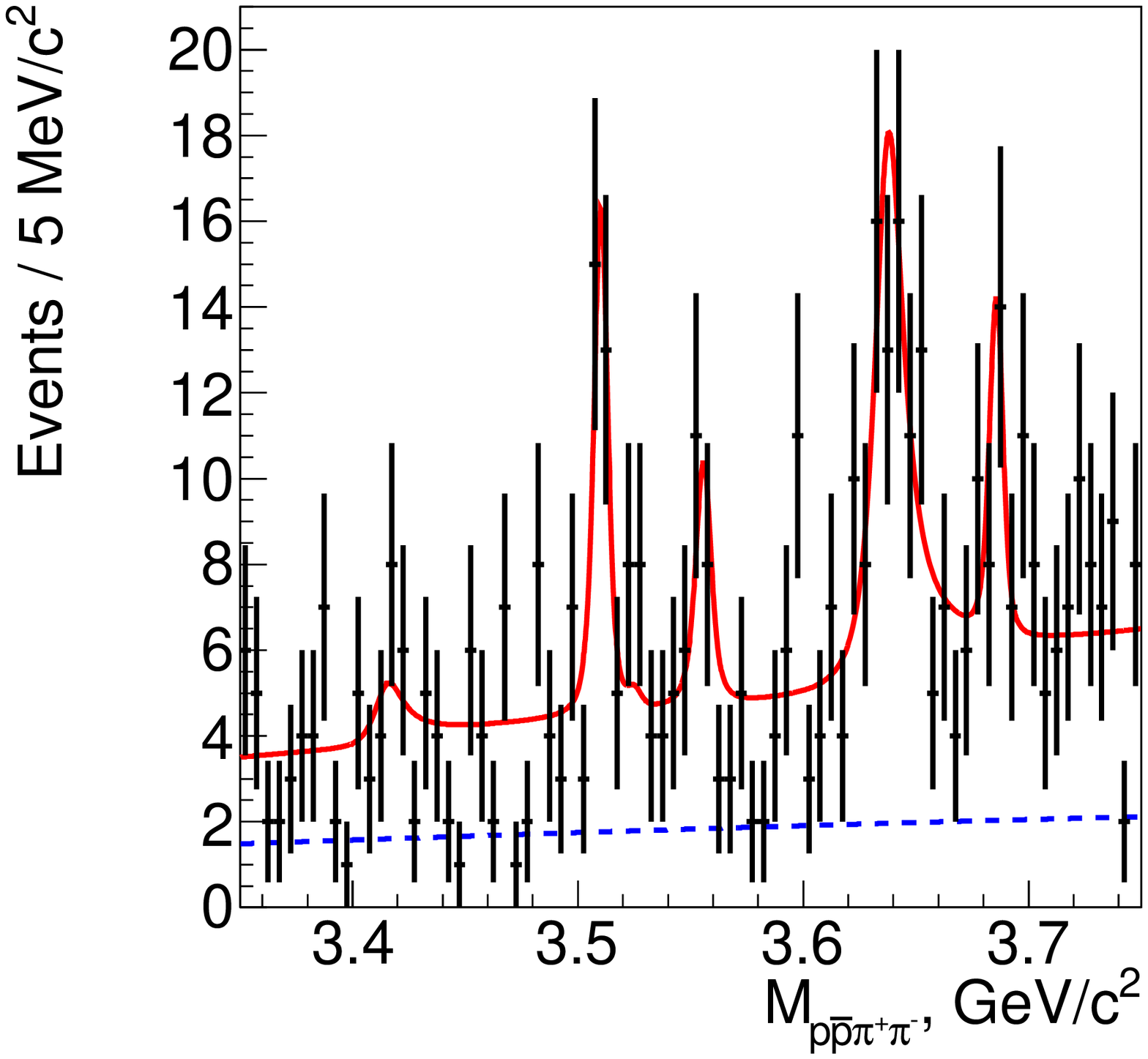} \\
\caption{Fit results in the $\decayks$ channel:
$\hcdec{0}$ signal (top left), $\hcdec{1}$ background (top right), 
$\hcdec{1}$ signal in the entire fitting region (bottom left)
and in the $\chi_{cJ}$ region (bottom right).
The red solid line is the fit result and the blue dashed line is the
background. The maximum of the $\hccha{1}$ signal fit result
at the $J/\psi$ peak is
more than two times greater than the number of data events in the corresponding
bin because the $J/\psi$ peak is narrower than the bin size. Thus,
a part of the fit result at the $J/\psi$ peak is not shown.}
\label{fig:hcks_fit_data}
\end{figure*}

The fit results are shown in Fig.~\ref{fig:hckp_fit_data} for the channel
$\decaykp$ and in Fig.~\ref{fig:hcks_fit_data} for the channel $\decayks$.
The signal yields and phases are listed in Table~\ref{tab:fit_data_parameters}.
The statistical significance of the decays $\decaykp$ and $\decayks$,
as well as the significances of other charmonium states in the channel
$\hccha{1}$ are calculated from the difference of $(-2 \ln L)$,
where $L$ is the maximum likelihood, between the
models with and without these states taking the number of degrees of freedom
into account. The significance of the decays $\decaykp$ and $\decayks$
in the default model is found to be $5.0\sigma$ and $0.8\sigma$, respectively.
The significance of the decays $\decaykp$ and $\decayks$ with
the systematic error taken into account is $4.8\sigma$ and $0.7\sigma$,
respectively; the procedure of the calculation of the systematic uncertainty
is described in Sec.~\ref{sec:systematic}.
Thus, we find evidence for the decay $\decaykp$,
but do not find evidence for $\decayks$.
The significances of charmonium states in the channel $\hccha{1}$
(except the $h_c$, which is reconstructed in two decay channels)
are shown in Table~\ref{tab:significance_pppipi}.
The significance of the $\eta_c(2S)$
in the default model is $12.3\sigma$ and $5.9\sigma$
for the processes $\decayccchakp{1}$ and $\decayccchaks{1}$, respectively.
The significance including the systematic error is $12.1\sigma$ and
$5.8\sigma$, respectively. Consequently, the decay $\eta_c(2S) \to \hccha{1}$
is observed for the first time in both $\decayccchakp{1}$ and
$\decayccchaks{1}$ processes.

\begin{table}
\caption{The resulting signal yields and phases in the default model.
The errors are statistical only.}
\begin{tabular}{c|c|c}
\hline\hline
Parameter & $\decaykp$ & $\decayks$ \\
\hline
$N_{\eta_c}$ & $229 \pm 18$ & $96 \pm 11$ \\
$\varphi_{\eta_c}$ & $-1.12 \pm 0.09$ & $-1.47 \pm 0.14$ \\
$N_{J/\psi}$ & $345 \pm 19$ & $128 \pm 12$ \\
$N_{\chi_{c0}}$ & $25.5 \pm 7.1$ & $0.9 \pm 1.6$ \\
$\varphi_{\chi_{c0}}$ & $-1.51 \pm 0.23$ & $-2.05 \pm 1.20$ \\
$N_{\chi_{c1}}$ & $34.5 \pm 8.8$ & $21.2 \pm 6.1$ \\
$N_{h_c}$ & $32.6 \pm 8.0$ & $3.1 \pm 3.8$ \\
$N_{\chi_{c2}}$ & $-1.6 \pm 6.3$ & $11.6 \pm 5.5$ \\
$N_{\eta_c(2S)}$ & $86.1 \pm 11.9$ & $24.0 \pm 6.8$ \\
$\varphi_{\eta_c(2S)}$ & $-1.41 \pm 0.14$ & $-1.90 \pm 0.23$ \\
$N_{\psi(2S)}$ & $36.9 \pm 8.8$ & $13.0 \pm 5.6$ \\
\hline\hline
\end{tabular}
\label{tab:fit_data_parameters}
\end{table}

\begin{table}
\caption{Significances of the charmonium states decaying to $\hccha{1}$
except the $h_c$ in the default model.}
\begin{tabular}{c|c|c}
\hline\hline
State & $\Bp \to (c\bar{c}) \kp$ & $\Bz \to (c\bar{c}) \ks$ \\
\hline
$\eta_c$     & $20.1\sigma$ & $12.5\sigma$ \\
$J/\psi$     & $33.9\sigma$ & $20.8\sigma$ \\
$\chi_{c0}$  &  $6.0\sigma$ &  $0.6\sigma$ \\
$\chi_{c1}$  &  $4.9\sigma$ &  $4.5\sigma$ \\
$\chi_{c2}$  &  $0.3\sigma$ &  $2.5\sigma$ \\
$\eta_c(2S)$ & $12.3\sigma$ &  $5.9\sigma$ \\
$\psi(2S)$   &  $5.0\sigma$ &  $2.8\sigma$ \\
\hline\hline
\end{tabular}
\label{tab:significance_pppipi}
\end{table}

\subsection{Systematic uncertainty: model dependence}
\label{sec:systematic}

For a systematic-uncertainty study, we consider additional models. They include
the models with free masses and widths of the $h_c$ and
all other charmonium states (with Gaussian constraints in
accordance with the errors of their current world average values), with
increased order (3) of the background PDF polynomial, different
fitting ranges, scaled resolution, and variation of
the relative fraction of the channels $\hcdec{0}$ and $\hcdec{1}$.
For the model with scaled resolution, the resolution function
$R_{h_c}(\Delta M)$ is changed to
\begin{equation}
R_{h_c}(\Delta M) \to \frac{1}{\mathcal{S}}
R_{h_c}\left(\frac{\Delta M}{\mathcal{S}}\right),
\end{equation}
where $\mathcal{S}$ is the resolution scaling parameter.
The variation of
the relative fraction of the channels $\hcdec{0}$ and $\hcdec{1}$
is performed by changing the expected yields in these channels by
$\pm 1 \sigma$, where the error is due to the error of the corresponding
branching fractions.
The results are listed in Table~\ref{tab:hc_significance_systematics}.

\begin{table*}
\caption{Model dependence of the $h_c$ and $\eta_c(2S) \to \hccha{1}$
significance.}
\begin{tabular}{c|c|c|c|c}
\hline\hline
\multirow{2}{*}{Model} & \multicolumn{2}{|c|}{$h_c$ significance} &
\multicolumn{2}{|c}{$\eta_c(2S) \to \hccha{1}$ significance}\\
\cline{2-5}
& $\decaykp$ & $\decayks$ & $\decaykp$ & $\decayks$ \\
\hline
Default                                   &
$5.0\sigma$ & $0.8\sigma$ & $12.3\sigma$ & $5.9\sigma$ \\
Free masses and widths                    &
$5.0\sigma$ & $0.8\sigma$ & $12.3\sigma$ & $6.0\sigma$ \\
Polynomial order ($\hcdec{0}$)            &
$4.8\sigma$ & $0.8\sigma$ & $12.3\sigma$ & $5.9\sigma$ \\
Polynomial order ($\hcdec{1}$ background) &
$5.0\sigma$ & $0.8\sigma$ & $12.2\sigma$ & $5.9\sigma$ \\
Polynomial order ($\hcdec{1}$ signal)     &
$5.0\sigma$ & $0.9\sigma$ & $12.2\sigma$ & $5.9\sigma$ \\
Fitting range variation ($\hcdec{0}$)     &
$5.0\sigma$ & $0.9\sigma$ & $12.3\sigma$ & $5.9\sigma$ \\
Fitting range variation ($\hcdec{1}$)     &
$5.0\sigma$ & $0.8\sigma$ & $12.1\sigma$ & $5.8\sigma$ \\
Scaled resolution                         &
$5.0\sigma$ & $0.8\sigma$ & $12.3\sigma$ & $6.0\sigma$ \\
Fraction of $\hcdec{0}$ and $\hcdec{1}$   &
$4.9\sigma$ & $0.7\sigma$ & $12.2\sigma$ & $5.9\sigma$ \\
\hline\hline
\end{tabular}
\label{tab:hc_significance_systematics}
\end{table*}

\subsection{Branching fraction}

Using the number of reconstructed events, we calculate the branching fractions
of the decays $\decaykp$ and $\decayks$, as well as the branching fraction
products $\br(\decaycckp) \times \br((c\bar{c}) \to \hccha{1})$
and $\br(\decayccks) \times \br((c\bar{c}) \to \hccha{1})$.
The decay $\psi(2S) \to \hccha{1}$ can proceed via the $J/\psi$:
$\psi(2S) \to J/\psi (\to p \bar{p}) \pip \pim$. To remove the events with
a $J/\psi$, the $\psi(2S)$ yield is taken from an alternative fit with an
additional $J/\psi$ veto defined as
$|M_{p \bar{p}} - m_{J/\psi}| > 50\ \mevcc$.

The sources of the systematic uncertainty of the branching fractions include
the model dependence
(the same set of alternative fit models is used as in
Sec.~\ref{sec:systematic}),
overtraining (the difference between the efficiency in the training and testing
samples),
the error of the difference of the particle identification requirements
efficiency between the data and MC,
the difference of the MLP efficiency between the data and MC,
tracking efficiency, number of $\Upsilon(4S)$ events,
the $\etacdec{0}$ and
$\Upsilon(4S) \to B^+ B^-$ or $\Upsilon(4S) \to B^0 \bar{B}^0$
branching fractions.
All systematic error sources are listed in
Table~\ref{tab:hckp_branching_systematic_error} for the channel
$\decaycckp$ ($\decaykp$ for the $h_c$ and
$\decayccchakp{1}$ for all other charmonium states) and in
Table~\ref{tab:hcks_branching_systematic_error}
for the channel $\decayccks$ ($\decayks$ for the $h_c$ and
$\decayccchaks{1}$ for all other charmonium states).
The errors of the tracking efficiency and the difference of
the efficiency of the particle identification requirements
depend on the multivariate-analysis channel in case of the calculation of
$\br(\decaykp)$ and $\br(\decayks)$;
the errors related to the MLP efficiency and overtraining
are estimated separately for $\hcdec{0}$ and $\hcdec{1}$.
The values presented in
Tables~\ref{tab:hckp_branching_systematic_error}
and \ref{tab:hcks_branching_systematic_error}
are weighted averages.

\begin{table*}
\caption{Relative systematic uncertainties of the branching fractions
for the channel $\decaycckp$ ($\decaykp$ for the $h_c$ and
$\decayccchakp{1}$ for all other charmonium states).}
\begin{tabular}{c|c|c|c|c|c|c|c|c}
\hline\hline
Error source & $h_c$ & $\eta_c$ & $J/\psi$ & $\chi_{c0}$ & $\chi_{c1}$ & $\chi_{c2}$ & $\eta_c(2S)$ & $\psi(2S)$ \\
\hline
Model dependence & $(\err{9.09}{9.16})\%$ & $(\err{3.45}{1.47})\%$ & $(\err{1.96}{0.04})\%$ & $(\err{4.59}{5.32})\%$ & $(\err{7.10}{2.98})\%$ & $(\err{21.13}{65.93})\%$ & $(\err{1.75}{4.62})\%$ & $(\err{1.74}{7.07})\%$ \\
PID & 3.99\% & 3.64\% & 3.62\% & 3.50\% & 3.50\% & 3.51\% & 3.52\% & 3.53\% \\
Overtraining & 0.41\% & \multicolumn{7}{|c}{0.14\%} \\
Tracking & 1.60\% & \multicolumn{7}{|c}{1.75\%} \\
MLP efficiency & 12.73\% & \multicolumn{7}{|c}{0.25\%} \\
Number of $\piz$ candidates & 11.60\% & \multicolumn{7}{|c}{---} \\
$\eta_c$ mass and width & 0.99\% & \multicolumn{7}{|c}{---} \\
$h_c$ branching fraction & 10.22\% & \multicolumn{7}{|c}{---} \\
$\br(\Upsilon(4S) \to B^+ B^-)$ & \multicolumn{8}{|c}{1.17\%} \\
Number of $\Upsilon(4S)$ events & \multicolumn{8}{|c}{1.37\%} \\
\hline
Total & $(\err{22.51}{22.54})\%$ & $(\err{5.62}{4.67})\%$ & $(\err{4.83}{4.41})\%$ & $(\err{6.30}{6.86})\%$ & $(\err{8.31}{5.25})\%$ & $(\err{21.57}{66.07})\%$ & $(\err{4.68}{6.34})\%$ & $(\err{4.68}{8.29})\%$ \\
\hline\hline
\end{tabular}
\label{tab:hckp_branching_systematic_error}
\end{table*}

\begin{table*}
\caption{Relative systematic uncertainties of the branching fractions
for the channel $\decayccks$ ($\decayks$ for the $h_c$ and
$\decayccchaks{1}$ for all other charmonium states).}
\begin{tabular}{c|c|c|c|c|c|c|c|c}
\hline\hline
Error source & $h_c$ & $\eta_c$ & $J/\psi$ & $\chi_{c0}$ & $\chi_{c1}$ & $\chi_{c2}$ & $\eta_c(2S)$ & $\psi(2S)$ \\
\hline
Model dependence & $(\err{30.94}{24.05})\%$ & $(\err{4.91}{24.37})\%$ & $(\err{1.45}{1.80})\%$ & $(\err{134.01}{24.56})\%$ & $(\err{6.79}{2.28})\%$ & $(\err{8.63}{4.78})\%$ & $(\err{4.51}{5.36})\%$ & $(\err{10.89}{9.01})\%$ \\
PID & 4.86\% & 3.93\% & 3.93\% & 3.87\% & 3.87\% & 3.86\% & 3.83\% & 3.81\% \\
Overtraining & 0.15\% & \multicolumn{7}{|c}{0.19\%} \\
Tracking & 1.95\% & \multicolumn{7}{|c}{2.10\%} \\
MLP efficiency & 12.79\% & \multicolumn{7}{|c}{0.25\%} \\
Number of $\piz$ candidates & 11.66\% & \multicolumn{7}{|c}{---} \\
$\eta_c$ mass and width & 0.96\% & \multicolumn{7}{|c}{---} \\
$h_c$ branching fraction & 10.27\% & \multicolumn{7}{|c}{---} \\
$\br(\Upsilon(4S) \to B^0 \bar{B}^0)$ & \multicolumn{8}{|c}{1.23\%} \\
Number of $\Upsilon(4S)$ events & \multicolumn{8}{|c}{1.37\%} \\
\hline
Total & $(\err{37.33}{31.86})\%$ & $(\err{6.89}{24.85})\%$ & $(\err{5.04}{5.16})\%$ & $(\err{134.10}{25.02})\%$ & $(\err{8.31}{5.30})\%$ & $(\err{9.87}{6.76})\%$ & $(\err{6.56}{7.17})\%$ & $(\err{11.88}{10.18})\%$ \\
\hline\hline
\end{tabular}
\label{tab:hcks_branching_systematic_error}
\end{table*}

The difference of the particle identification requirements
efficiency between the data and MC is estimated from several control samples,
such as $D^{*+}\to D^0(\to K^-\pi^+)\pi^+$ for $K$ and $\pi$,
$\Lambda \to p \pip$ for $p$, $\Lambda_c^+ \to \Lambda \pip$ for $\Lambda$,
$\Dz \to \ks \pip \pim$ for $\ks$, and $\tau^- \to \pim \piz \nu_\tau$ for
$\piz$. The resulting overall efficiency ratio depends on the final state and
the momenta of the decay products; thus, it is different for all branching
fractions.
For example, for the $h_c$ it is found to be
$(95.0 \pm 3.8)\%$ for the channel $\decaykp$ and $(93.0 \pm 4.5)\%$ for the
channel $\decayks$.

The error caused by the difference of the MLP efficiency between the data and
MC is estimated for the channel $\hcdec{0}$
using the decay mode $\Bz \to \eta_c \pim \kp$. This decay is
reconstructed using selection criteria that are as similar as possible to the
signal mode $\decaykp$. The same MLP optimized for $\decaykp$ is applied to
the control channel. Some MLP input variables used for the signal channel are
undefined for $\Bz \to \eta_c \pim \kp$, for example, the number of $\piz$
candidates that include the $h_c$ daughter $\gamma$ as one of their daughters.
Such variables are held constant. The ratio of the
number of signal candidates before and after the application of the MLP
selection requirements is used to measure the difference of the MLP
efficiency between the data and MC:
\begin{equation}
r_\text{MLP} = \frac{N(\text{all $\eta_c$ channels, with MLP cut})}
{N(\etacdec{0},\ \text{no MLP cut})}.
\end{equation}
Only the channel $\etacdec{0}$ is used before the MLP selection,
because only this channel is sufficiently clean for the determination of
the number of the signal events without the MLP selection.
The ratio $r_\text{MLP}$ is extracted from a simultaneous fit to the
$\eta_c$ mass distribution before and after the application of the MLP
selection.
The relative difference between the values of $r_\text{MLP}$ in data
and MC is found to be $(14.4 \pm 9.0)\%$.
For conservative treatment, the statistical error is added in quadrature
to the central value of the difference.
The resulting systematic uncertainty caused by the MLP selection
efficiency difference in data and MC is 17.0\%.

The estimation of the MLP efficiency error for the hadronic
channel $\hccha{1}$ is done by performing the fit using the $\hccha{1}$
data without the MLP selection and comparing the resulting branching fraction
products $\br(\Bp \to J/\psi \kp) \times \br(J/\psi \to \hccha{1})$ with the
results of the default procedure.
Their relative difference is found to be $0.2\%$.

The same estimates of the MLP efficiency uncertainty for the channels
$\hcdec{0}$ and $\hcdec{1}$ are used for both $\decaykp$ and $\decayks$.
The final value of the MLP efficiency uncertainty is calculated
as a weighted average of the errors for the two $h_c$ decay channels.
The result is slightly different for the channels $\decaykp$ and $\decayks$
because of the difference in the relative number of the expected
$\hcdec{0}$ and $\hcdec{1}$ events.
The MLP efficiency error for the branching fraction products for the
channels $\Bp \to (c\bar{c}) (\to \hccha{1}) \kp$ and
$\Bz \to (c\bar{c}) (\to \hccha{1}) \ks$ is equal to the error for the
channel $\hccha{1}$, since all charmonium states other than the $h_c$ are
reconstructed in this channel only.

\begin{table*}
\caption{Measured branching fractions and branching fraction products
and their comparison with the current world-average
values~\cite{Tanabashi:2018oca}.}
\begin{tabular}{c|c|c}
\hline\hline
Branching fraction & Value or confidence interval (90 \% C. L.) & World-average value \\
\hline
$\br(\Bp \to h_c \kp)$ & $(3.7 \err{1.0}{0.9} \err{0.8}{0.8}) \times 10^{-5}$ & $< 3.8 \times 10^{-5}$ \\
$\br(\Bp \to \eta_c \kp) \times \br(\eta_c \to \hccha{1})$ & $(39.4 \err{4.1}{3.9} \err{2.2}{1.8}) \times 10^{-7}$ & $(57.8 \pm 20.2) \times 10^{-7}$ \\
$\br(\Bp \to J/\psi \kp) \times \br(J/\psi \to \hccha{1})$ & $(56.4 \err{3.3}{3.2} \err{2.7}{2.5}) \times 10^{-7}$ & $(60.6 \pm 5.3) \times 10^{-7}$\\
$\br(\Bp \to \chi_{c0} \kp) \times \br(\chi_{c0} \to \hccha{1})$ & $(3.7 \err{1.2}{1.0} \err{0.2}{0.3}) \times 10^{-7}$ & $(3.1 \pm 1.1) \times 10^{-7}$ \\
$\br(\Bp \to \chi_{c1} \kp) \times \br(\chi_{c1} \to \hccha{1})$ & $(4.7 \err{1.3}{1.2} \err{0.4}{0.2}) \times 10^{-7}$ & $(2.4 \pm 0.9) \times 10^{-7}$ \\
$\br(\Bp \to \chi_{c2} \kp) \times \br(\chi_{c2} \to \hccha{1})$ & $< 1.9 \times 10^{-7}$ & $(0.15 \pm 0.06) \times 10^{-7}$ \\
$\br(\Bp \to \eta_c(2S) \kp) \times \br(\eta_c(2S) \to \hccha{1})$ & $(11.2 \err{1.8}{1.6} \err{0.5}{0.7}) \times 10^{-7}$ & not seen \\
$\br(\Bp \to \psi(2S) \kp) \times \br(\psi(2S) \to \hccha{1})$ & $[0.5, 3.5] \times 10^{-7}$ & $(3.7 \pm 0.3) \times 10^{-7}$ \\
\hline
$\br(\Bz \to h_c \ks)$ & $< 1.4 \times 10^{-5}$ & not seen \\
$\br(\Bz \to \eta_c \ks) \times \br(\eta_c \to \hccha{1})$ & $(19.0 \err{3.2}{2.9} \err{1.3}{4.7}) \times 10^{-7}$ & $(20.9 \pm 7.8) \times 10^{-7}$ \\
$\br(\Bz \to J/\psi \ks) \times \br(J/\psi \to \hccha{1})$ & $(24.3 \err{2.3}{2.2} \err{1.2}{1.3}) \times 10^{-7}$ & $(26.2 \pm 2.4) \times 10^{-7}$ \\
$\br(\Bz \to \chi_{c0} \ks) \times \br(\chi_{c0} \to \hccha{1})$ & $< 1.3 \times 10^{-7}$ & $(1.5 \pm 0.6) \times 10^{-7}$ \\
$\br(\Bz \to \chi_{c1} \ks) \times \br(\chi_{c1} \to \hccha{1})$ & $(3.7 \err{1.2}{1.0} \err{0.3}{0.2}) \times 10^{-7}$ & $(1.0 \pm 0.4) \times 10^{-7}$ \\
$\br(\Bz \to \chi_{c2} \ks) \times \br(\chi_{c2} \to \hccha{1})$ & $[0.7, 3.8] \times 10^{-7}$ & not seen \\
$\br(\Bz \to \eta_c(2S) \ks) \times \br(\eta_c(2S) \to \hccha{1})$ & $(4.2 \err{1.4}{1.2} \err{0.3}{0.3}) \times 10^{-7}$ & not seen \\
$\br(\Bz \to \psi(2S) \ks) \times \br(\psi(2S) \to \hccha{1})$ & $< 1.9 \times 10^{-7}$ & $(1.7 \pm 0.2) \times 10^{-7}$ \\
\hline\hline
\end{tabular}
\label{tab:branching}
\end{table*}

The MLP efficiency uncertainty for the channel $\hcdec{0}$ does not include
the uncertainty caused by the difference between the data and MC in
the distributions of the variables that are not defined for the channel
$\Bz \to \eta_c \pim \kp$. There are four such variables: the $\eta_c$ mass,
the $h_c$ helicity angle, and two numbers of $\piz$ candidates that include
the $h_c$ daughter photon as one of their daughters. The distribution of the
$h_c$ helicity angle for the signal events is known precisely; thus, there is
no additional uncertainty caused by the difference of its distribution in
data and MC. The difference of the numbers of $\piz$ candidates is taken
into account by removing these variables from the neural network for the
channel $\decaykp$, performing an alternative optimization, and comparing the
resulting $h_c$ branching fractions in the channel $\hcdec{0}$. The relative
difference is found to be $15.6\%$. The error due to the $\eta_c$ mass
distribution uncertainty is estimated by varying the $\eta_c$ mass and width
by $\pm 1 \sigma$ and reweighting the selected MC events in accordance with
the relative difference between the modified and default $\eta_c$ mass
distributions. The largest
resulting efficiency difference is considered as the systematic uncertainty
related to the $\eta_c$ mass distribution. This uncertainty is estimated to be
$1.3\%$ for both $\decaykp$ and $\decayks$ channels.

The ratio $r_\text{MLP}$ used for determination of the MLP efficiency
uncertainty includes the number of reconstructed events relatively to the
number of events in $\optdec{0}$. The total expected number of events
can be calculated as
\begin{equation}
2 N_{\Upsilon(4S)} \left(\sum\limits_i \epsilon_i \br_i\right) =
2 N_{\Upsilon(4S)} \br_{\hcdec{0}} \left(\sum\limits_i \epsilon_i
\frac{\br_i}{\br_{\hcdec{0}}}\right),
\label{eq:branching_efficiency}
\end{equation}
where $\epsilon_i$ and $\br_i$
are the efficiency and branching fraction for $i$-th channel, respectively.
The last term in Eq.~\eqref{eq:branching_efficiency} is proportional to
$r_\text{MLP}$. Consequently, for the channel $\hcdec{0}$ one needs to
take into account only the error of $\br(\optdec{0})$. Errors of
the branching fractions of all other channels relatively to $\optdec{0}$
enter the MLP efficiency error.

The final $h_c$ branching fraction error is calculated as a weighted average
of the errors for the channels $\hcdec{0}$ and $\hcdec{1}$. Since branching
fraction products are measured for all other charmonium states, they do not
have a similar systematic error source.

The resulting branching fractions with both statistical and systematic
errors are listed in Table~\ref{tab:branching}. 
For insignificant decays or decay chains, the
confidence intervals are calculated
in the frequentist approach~\cite{Feldman:1997qc} using an asymmetric Gaussian
as the branching-fraction PDF and the measured central value
and errors as its parameters. This is the first measurement of
$\br(\decayks)$. Also, the branching fraction products
$\br(\decaycckp) \times \br((c\bar{c}) \to \hccha{1})$
and $\br(\decayccks) \times \br((c\bar{c}) \to \hccha{1})$
are measured directly in $B$ decays for the first time.
The current world-average values of the same branching
fractions~\cite{Tanabashi:2018oca}
are also presented in Table~\ref{tab:branching} for comparison if they are
known. The values of the branching fraction products are calculated by
multiplying the individual branching fractions listed in
Ref.~\cite{Tanabashi:2018oca} assuming uncorrelated errors.
The measured branching fractions are consistent with the world averages;
the largest deviation is observed for the branching-fraction
product $\br(\Bz \to \chi_{c1} \ks) \times \br(\chi_{c1} \to \hccha{1})$,
which differs from the world-average value by $2.4\sigma$ taking its error
into account.
The results for $\br(\Bp \to \eta_c \kp) \times \br(\eta_c \to \hccha{1})$
and $\br(\Bz \to \eta_c \ks) \times \br(\eta_c \to \hccha{1})$ have a better
precision than the world-average values.

\section{Conclusions}

A search for the decays $\decaykp$ and $\decayks$ has been performed.
Evidence for the decay $\decaykp$ is found; its
significance is $4.8\sigma$.
No evidence is found for $\decayks$.
The branching fraction of $\decaykp$ is measured to be
$(3.7 \err{1.0}{0.9} \err{0.8}{0.8}) \times 10^{-5}$; the upper limit for
the $\decayks$ branching fraction is $1.4 \times 10^{-5}$ at $90\%$ C. L.
The measured value of $\br(\decaykp)$ is consistent with the existing upper
limit of $3.8 \times 10^{-5}$ ($90\%$ C. L.) obtained in the previous Belle
analysis~\cite{Fang:2006bz} and supersedes it. The resulting branching fraction
$\br(\decaykp)$ agrees with the existing theoretical
predictions~\cite{Meng:2006mi,Li:2006vj,Beneke:2008pi}.
In addition, a study of the $\hccha{1}$ invariant mass
distribution in the channel $\Bp \to (\hccha{1}) \kp$ results in the first
observation of the decay $\eta_c(2S) \to \hccha{1}$ with
$12.1\sigma$ significance.

\section*{Acknowledgement}

We thank the KEKB group for the excellent operation of the
accelerator; the KEK cryogenics group for the efficient
operation of the solenoid; and the KEK computer group, and the Pacific Northwest National
Laboratory (PNNL) Environmental Molecular Sciences Laboratory (EMSL)
computing group for strong computing support; and the National
Institute of Informatics, and Science Information NETwork 5 (SINET5) for
valuable network support.  We acknowledge support from
the Ministry of Education, Culture, Sports, Science, and
Technology (MEXT) of Japan, the Japan Society for the 
Promotion of Science (JSPS), and the Tau-Lepton Physics 
Research Center of Nagoya University; 
the Australian Research Council including grants
DP180102629, 
DP170102389, 
DP170102204, 
DP150103061, 
FT130100303; 
Austrian Science Fund (FWF);
the National Natural Science Foundation of China under Contracts
No.~11435013,  
No.~11475187,  
No.~11521505,  
No.~11575017,  
No.~11675166,  
No.~11705209;  
Key Research Program of Frontier Sciences, Chinese Academy of Sciences (CAS), Grant No.~QYZDJ-SSW-SLH011; 
the  CAS Center for Excellence in Particle Physics (CCEPP); 
the Shanghai Pujiang Program under Grant No.~18PJ1401000;  
the Ministry of Education, Youth and Sports of the Czech
Republic under Contract No.~LTT17020;
the Carl Zeiss Foundation, the Deutsche Forschungsgemeinschaft, the
Excellence Cluster Universe, and the VolkswagenStiftung;
the Department of Science and Technology of India; 
the Istituto Nazionale di Fisica Nucleare of Italy; 
National Research Foundation (NRF) of Korea Grants
No.~2015H1A2A1033649, No.~2016R1D1A1B01010135, No.~2016K1A3A7A09005
603, No.~2016R1D1A1B02012900, No.~2018R1A2B3003 643,
No.~2018R1A6A1A06024970, No.~2018R1D1 A1B07047294; Radiation Science Research Institute, Foreign Large-size Research Facility Application Supporting project, the Global Science Experimental Data Hub Center of the Korea Institute of Science and Technology Information and KREONET/GLORIAD;
the Polish Ministry of Science and Higher Education and 
the National Science Center;
the Russian Foundation for Basic Research Grant No.~18-32-00277;
the Slovenian Research Agency;
Ikerbasque, Basque Foundation for Science, Spain;
the Swiss National Science Foundation; 
the Ministry of Education and the Ministry of Science and Technology of Taiwan;
and the United States Department of Energy and the National Science Foundation.


\begin{thebibliography}{99}

\bibitem{Bauer:1986bm} 
  M.~Bauer, B.~Stech and M.~Wirbel,
  Z.\ Phys.\ C {\bf 34}, 103 (1987).

\bibitem{Suzuki:2002sq} 
  M.~Suzuki,
  Phys.\ Rev.\ D {\bf 66}, 037503 (2002).

\bibitem{Tanabashi:2018oca} 
  M.~Tanabashi {\it et al.} (Particle Data Group),
  Phys.\ Rev.\ D {\bf 98}, 030001 (2018).

\bibitem{Fang:2006bz} 
  F.~Fang {\it et al.} (Belle Collaboration),
  Phys.\ Rev.\ D {\bf 74}, 012007 (2006).

\bibitem{Aubert:2008kp} 
  B.~Aubert {\it et al.} (BaBar Collaboration),
  Phys.\ Rev.\ D {\bf 78}, 012006 (2008).

\bibitem{Aaij:2013rha} 
  R.~Aaij {\it et al.} (LHCb Collaboration),
  Eur.\ Phys.\ J.\ C {\bf 73}, 2462 (2013).

\bibitem{Aaij:2016kxn}
  R.~Aaij {\it et al.} (LHCb Collaboration),
  Phys.\ Lett.\ B {\bf 769}, 305 (2017).

\bibitem{Meng:2006mi} 
  C.~Meng, Y.~J.~Gao and K.~T.~Chao,
  hep-ph/0607221.

\bibitem{Li:2006vj} 
  X.~Q.~Li, X.~Liu and Y.~M.~Wang,
  Phys.\ Rev.\ D {\bf 74}, 114029 (2006).

\bibitem{Beneke:2008pi} 
  M.~Beneke and L.~Vernazza,
  Nucl.\ Phys.\ B {\bf 811}, 155 (2009).

\bibitem{kekb}
{S.~Kurokawa and E.~Kikutani, Nucl. Instrum. Methods Phys. Res., Sect.
 A {\bf 499}, 1 (2003), and other papers included in this Volume;
 T.~Abe {\it et al.}, Prog. Theor. Exp. Phys. {\bf 2013}, 03A001 (2013)
 and references therein. }

\bibitem{Ablikim:2018ewr}
  M.~Ablikim {\it et al.} (BESIII Collaboration),
  Phys.\ Rev.\ D {\bf 99}, 072008 (2019).

\bibitem{Belle}
{A.~Abashian {\it et al.} (Belle Collaboration), Nucl. Instrum. Methods
 Phys. Res., Sect. A {\bf 479}, 117 (2002); also see detector section in
 J.~Brodzicka {\it et al.}, Prog. Theor. Exp. Phys. {\bf 2012}, 04D001 (2012). }

\bibitem{svd2}
  Z.~Natkaniec {\it et al.} (Belle SVD2 Group),
  Nucl. Instrum. Methods Phys. Res., Sect. A {\bf 560}, 1 (2006).

\bibitem{geant} R.~Brun \textit{et al.}, GEANT 3.21, CERN DD/EE/84-1, 1984.

\bibitem{evtgen}
  D.~J.~Lange,
  Nucl. Instrum. Methods Phys. Res., Sect. A {\bf 462}, 152 (2001).

\bibitem{Gelb:2018agf}
  M.~Gelb {\it et al.},
  Comput.\ Softw.\ Big Sci.\  {\bf 2}, 9 (2018).

\bibitem{Hanagaki:2001fz}
  K.~Hanagaki, H.~Kakuno, H.~Ikeda, T.~Iijima and T.~Tsukamoto,
  Nucl. Instrum. Methods Phys. Res., Sect. A {\bf 485}, 490 (2002).

\bibitem{Fox:1978vu} 
  G.~C.~Fox and S.~Wolfram,
  Phys.\ Rev.\ Lett.\ {\bf 41}, 1581 (1978).

\bibitem{Lees:2014iua}
  J.~P.~Lees {\it et al.} (BaBar Collaboration),
  Phys.\ Rev.\ D {\bf 89}, 112004 (2014).

\bibitem{skwarnicki}
T. Skwarnicki, Ph.D. Thesis, Institute for Nuclear Physics, Krakow 1986;
DESY Internal Report, DESY F31-86-02 (1986).

\bibitem{tmva}
  H.~Voss, A.~Hocker, J.~Stelzer and F.~Tegenfeldt,
  PoS ACAT, 040 (2007).

\bibitem{Punzi:2003bu}
  G.~Punzi,
  eConf C {\bf 030908}, MODT002 (2003).

\bibitem{Feldman:1997qc} 
  G.~J.~Feldman and R.~D.~Cousins,
  Phys.\ Rev.\ D {\bf 57}, 3873 (1998).

\end{thebibliography}
\end{document}